\documentclass[10pt,journal,twocolumn]{IEEEtran}
\usepackage{amssymb,latexsym,amsmath,amsfonts}
\usepackage{graphicx}
\usepackage{verbatim}
\usepackage{bm}			% allows us to import images
\newtheorem{theorem}{Theorem}
\newtheorem{lemma}{Lemma}
\newtheorem{remark}{Remark}
\newtheorem{proposition}{Proposition}
\begin{document}
\title{A Factor Graph Approach to Clock Offset Estimation in Wireless Sensor Networks}
%\title{Timing Synchronization in Wireless Sensor Networks for General Exponential Family Likelihoods}
\author{Aitzaz Ahmad, Davide Zennaro, Erchin Serpedin, and Lorenzo Vangelista
\thanks{A. Ahmad and E. Serpedin are with the Department
of Electrical and Computer Engineering, Texas A\&M University, Texas,
TX, 77843 USA.}
\thanks{D. Zennaro and L. Vangelista are with Department of Information Engineering, University of Padova, Padova, Italy.}
\thanks{A. Ahmad and E. Serpedin are supported by Qtel.}
\thanks{The work of D. Zennaro is partially supported by an ``A. Gini" fellowship and has been performed while on leave at Texas A\&M University, College Station, TX (USA).} }

\maketitle

\begin{abstract}
The problem of clock offset estimation in a two way timing message exchange regime is considered when the likelihood function of the observation time stamps is Gaussian, exponential or log-normally distributed. A parametrized solution to the maximum likelihood (ML) estimation of clock offset, based on convex optimization, is presented, which differs from the earlier approaches where the likelihood function is maximized graphically. In order to capture the imperfections in node oscillators, which may render a time-varying nature to the clock offset, a novel Bayesian approach to the clock offset estimation is proposed by using a factor graph representation of the posterior density. Message passing using the max-product algorithm yields a closed form expression for the Bayesian inference problem. Several lower bounds on the variance of an estimator are derived for arbitrary exponential family distributed likelihood functions which, while serving as stepping stones to benchmark the performance of the proposed clock offset estimators, can be useful in their own right in classical as well Bayesian parameter estimation theory. To corroborate the theoretical findings, extensive simulation results are discussed for classical as well as Bayesian estimators in various scenarios. It is observed that the performance of the proposed estimators is fairly close to the fundamental limits established by the lower bounds.

\end{abstract}


\section{Introduction}
\IEEEPARstart{W}{IRELESS} sensor networks (WSNs) typically consist of a large number of geographically distributed sensor nodes, deployed to observe some phenomenon of interest. The nodes constituting such a network are low cost sensors that have limited abilities of data processing and communication.
\begin{comment}The sensors report their observations to a distant fusion center which aggregates this data to infer the desired information. In addition, these sensor nodes can also collaborate to accomplish common targets. The on-board sensing equipment enables the sensors to summarize the useful information to be transmitted to the fusion center, resulting in reduced communication requirements.
\end{comment}
WSNs envisage tremendous applications in such diverse areas as industrial process control, battlefield surveillance, health monitoring, target localization and tracking, etc., \cite{akyildiz:survey}. With the recent
advances in digital circuit technology, WSNs are expected to play a pivotal role in future wireless communications.

Clock synchronization in sensor networks is a critical component in data fusion and duty cycling operations, and  has gained widespread interest in the past few years. Most of the current methods consider sensor networks exchanging time stamps based on the time at their respective clocks. A survey of the popular approaches employed in practice for timing synchronization is presented in \cite{sundar:survey} and \cite{sadler:survey}. The \emph{one-way} message exchange mechanism involves a reference node broadcasting its timing information to other nodes in a network. The receiver nodes record the arrival of these messages with respect to their own clock. After several such time stamps have been exchanged, the nodes estimate their offsets based on these observations. A particular case of this approach is the flooding time synchronization protocol (FTSP) \cite{maroti:ftsp} which uses regression to estimate the clock offset. On the other hand, through a \emph{two-way} timing exchange process, adjacent nodes aim to achieve pairwise synchronization by communicating their timing information with each other. After a round of $N$ messages, each nodes tries to estimate its own clock parameters. The timing-sync protocol for sensor networks (TPSNs) \cite{ganeriwal:tpsn} uses this strategy in two phases to synchronize clocks in a network. The level discovery phase involves a spanning tree based representation of a WSN while nodes attempt to synchronize with their immediate parents using a two-way message exchange process in the synchronization phase. In \emph{receiver-receiver} synchronization, nodes collect time stamps sent from a common broadcasting node and utilize them to adjust their clocks. The reference broadcast synchronization (RBS) protocol \cite{elson:rbs} uses reference beacons sent from a master node to establish a common notion of time across a network. An alternative framework for network-wide distributed clock synchronization consists of recasting the problem of agreement on oscillation phases and/or frequencies as a consensus based recursive model in which only local message passing is required among nodes. By assuming a connected network, it is possible to design efficient distributed algorithms by carefully choosing the update function. Under this framework, \cite{simeone:consensus} proposed a Laplacian-based algorithm for establishing agreement on oscillation frequencies all over the network based on standard consensus. A combined agreement over both clock phases and frequencies has been studied in \cite{zennaro:consensus}, by making use of state-of-the-art fast consensus techniques. Scalable synchronization algorithms for large sensor networks are developed in \cite{anna:2005} and \cite{anna:2011} inspired by mathematical biology models justifying synchrony in the biological agents.

The clock synchronization problem in a WSN offers a natural statistical signal processing framework whereby, the clock parameters are to be estimated using timing information from various sensors \cite{serpedin:survey}. A model based synchronization approach to arrest the clock drifts is explored in \cite{lemmon:model}. The impairments in message transmission arise from the various delays experienced by the messages as they travel through the transmission medium. Therefore, a crucial component of efficient clock parameter estimation is accurate modeling of the network delay distributions. Several distributions have been proposed that aim to capture the random queuing delays in a network \cite{bovy:end}. Some of these candidate distributions include exponential, Weibull, Gamma and log-normal distributions. Assuming an exponential delay distribution, several estimators were proposed in \cite{ghaffar:time-out}. It was argued that when the propagation delay $d$ is unknown, the maximum likelihood (ML) estimator for the clock offset $\theta$ is not unique. However, it was later shown in \cite{jeske:ML} that the ML estimator of $\theta$ does exist uniquely for the case of unknown $d$. The performance of these estimators was compared with benchmark estimation bounds in \cite{eddie:novel}. Considering an offset and skew model, Chaudhari \emph{et.al.} presented algorithms for the joint ML estimation of clock offset and skew in \cite{qasim:joint} when the network delays are exponentially distributed. Clock offset and skew estimators were determined in \cite{leng:gauss} based on the assumption that the network delays arise from the contribution of several independent processes and as such, were modeled as Gaussian. The convergence of  distributed consensus time synchronization algorithms is investigated in \cite{shalinee:consensus} and \cite{shalinee:secondorder}, assuming a Gaussian delay between sensor nodes. More recently, the minimum variance unbiased estimator (MVUE) for the clock offset under an exponential delay model was proposed in \cite{qasim:mvue}. The timing synchronization problem for the offset-only case was also recast as an instance of convex optimization in \cite{aitzaz:weibull} for Weibull distributed network delays. A recent contribution \cite{prkumar:2011} has investigated the feasibility of determining the clock parameters by studying the fundamental limits on clock synchronization for wireline and wireless networks.
\begin{comment}However, a common denominator in all the aforementioned approaches is the highly restrictive assumption on the network delay distribution. The estimators so obtained are not robust and their performance is severely degraded if the queuing delays do not conform with the assumptions. Clearly, a more general framework is required which works for several distributions and delivers performance with desired fidelity. It is precisely this avenue that we aim to investigate in this work.
\end{comment}

In this work, considering the classic two-way message exchange mechanism, a unified framework for the clock offset estimation problem is presented when the likelihood function of the observations is Gaussian, exponential or log-normally distributed. A parameterized solution is proposed for the ML estimation of clock offset by recasting the likelihood maximization as an instance of convex optimization. In order to incorporate the effect of time variations in the clock offset between sensor nodes, a Bayesian inference approach is also studied based on a factor graph representation of the posterior density. The major contributions of this work can be summarized as follows.
\begin{enumerate}
\item A unified framework for ML estimation of clock offset, based on convex optimization, is presented when the likelihood function of the observations is Gaussian, exponential and log-normally distributed. The proposed framework recovers the already known results for Gaussian and exponentially distributed likelihood functions and determines the ML estimate in case of log-normal distribution. Hence, the proposed convex optimization based approach represents a simpler alternative, and a more general derivation of ML estimator, which bypasses the graphical analysis used in \cite{jeske:ML} to maximize the likelihood function.
\item In order to capture the time variations in clock offsets due to imperfect oscillators, a Bayesian framework is presented by considering the clock offset as a random Gauss-Markov process. Bayesian inference is performed using factor graphs and the max-product algorithm. The message passing strategy yields a closed form solution for Gaussian, exponential and log-normally distributed likelihood functions. This extends the current literature to cases where the clock offset may not be deterministic, but is in fact a random process.
\item In order to evaluate the performance of the proposed estimators, classical as well as Bayesian bounds are derived for arbitrary exponential family distributed likelihood functions, which is a wide class and contains almost all distributions of interest. While these results aid in comparing various estimators in this work, they can be useful in their own right in classical and Bayesian estimation theory.
%\item Extensive simulation results are presented for various distributions e.g., Gaussian, exponential, log-normal and Gamma. These results corroborate the competitive performance of our approximate ML approach when the log-partition functions is considered a second degree polynomial in the parameter of interest.
\end{enumerate}
This paper is organized as follows. The system model is outlined in Section \ref{sec:model}. The ML estimation of clock offset based on convex optimization is proposed in Section \ref{sec:ml}. The factor graph based inference algorithm for the synchronization problem in a Bayesian paradigm is detailed in Section \ref{sec:factor} and a closed form solution is obtained. Section \ref{sec:bounds} presents several theoretical lower bounds on the variance of an estimator evaluated in the classical as well as Bayesian regime. Simulation studies are discussed in Section \ref{sec:simul} which corroborate the earlier results. Finally, the paper is concluded in Section \ref{sec:concl} along with some directions for future research.
%demo file is intended to serve as a ``starter file''
%for IEEE journal papers produced under \LaTeX\ using
%IEEEtran.cls version 1.7 and later.
%I wish you the best of success.

\section{System Model}
\begin{figure}[t]
\begin{center}
\includegraphics[scale=0.9]{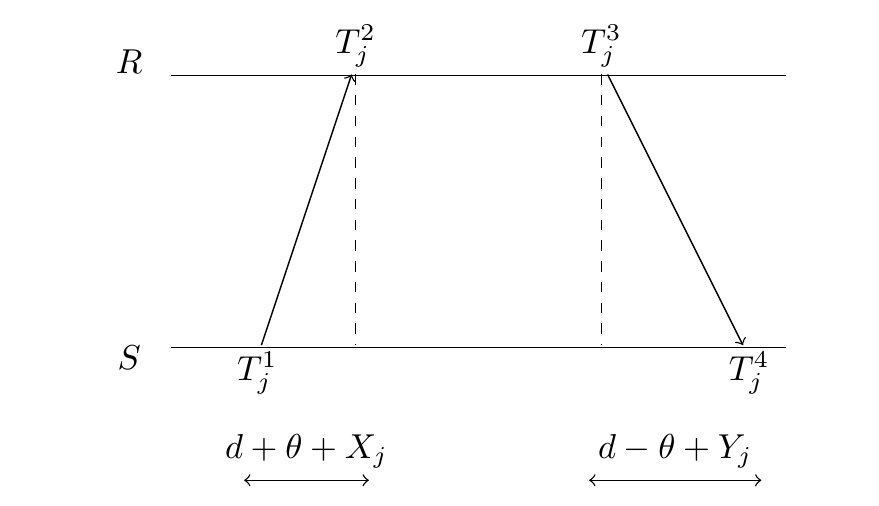}
\caption{A two-way timing message exchange mechanism}
\label{fig:time}
\end{center}
\end{figure}

The process of \label{sec:model}pairwise synchronization between two nodes $S$ and $R$ is illustrated in Fig. \ref{fig:time}. At the $j$th message exchange, node $S$ sends the information about its current time through a message including time stamp $T_j^1$. Upon receipt of this message, Node $R$ records the reception time $T_j^2$ according to its own time scale. The two-way timing message exchange process is completed when node $R$ replies with a synchronization packet containing time stamps $T_j^2$ and $T_j^3$ which is received at time $T_j^4$ by node $S$ with respect to its own clock. After $N$ such messages have been exchanged between nodes $S$ and $R$, node $S$ is equipped with time stamps $\{T_j^1, T_j^2, T_j^3, T_j^4\}_{j=1}^{N}$. The impairments in the signaling mechanism occur due to a \emph{fixed} propagation delay, which accounts for the time required by the message to travel through the transmission medium, and a \emph{variable} network delay, that arises due to queuing delay experienced by the messages during transmission and reception. By assuming that the respective clocks of nodes $S$ and $R$ are related by $C_R(t)=\theta + C_S(t)$, the two-way timing message exchange model at the $j$th instant can be represented as
\begin{align}
T_j^2&=T_j^1+d+\theta+X_j\nonumber \\
T_j^4&=T_j^3+d-\theta+Y_j\label{time:stamp}
\end{align}
where $d$ represents the propagation delay, assumed symmetric in both directions, and $\theta$ is offset of the clock at node $R$ relative to the clock at node $S$. $X_j$ and $Y_j$ are the independent and identically distributed variable network delays. By defining \cite{ghaffar:time-out}
\begin{align*}
U_j&\overset{\Delta}{=}T_j^2-T_j^1\\
V_j&\overset{\Delta}{=}T_j^4-T_j^3 \;,
\end{align*}
the system in \eqref{time:stamp} can be equivalently expressed as
\begin{align}
U_j&=d+\theta+X_j\nonumber \\
V_j&=d-\theta+Y_j\label{U:V}\;.
\end{align}
By further defining
\begin{equation}
\xi\overset{\Delta}{=}d+\theta \qquad
\psi\overset{\Delta}{=}d-\theta \label{subs}\;,
\end{equation}
the model in \eqref{U:V} can be written as
\begin{align*}
U_j&=\xi+X_j\nonumber \\
V_j&=\psi+Y_j
\end{align*}
for $j=1,\ldots,N$.
\begin{comment}
The ML estimates of the parameters $d$ and $\theta$ can be equivalently determined from the ML estimates of $\xi$ and $\psi$ using the invariance principle \cite{kay:est}.
\end{comment}
The goal is to determine precise estimates of $\xi$ and $\psi$ using observations $\{U_j, V_j\}_{j=1}^{N}$. An estimate of $\theta$ can, in turn, be obtained using \eqref{subs} as follows
\begin{equation}
\theta = \frac{\xi-\psi}{2}\label{subs:theta}\;.
\end{equation}

Accurate modeling of the variable delays, $X_j$ and $Y_j$, has been a topic of interest in recent years. Several distributions have been proposed that aim to capture the random effects caused by the queuing delays \cite{bovy:end}. These distributions include exponential, gamma, log-normal and Weibull. In addition, the authors in \cite{leng:gauss} argued that $X_j$ and $Y_j$ result from contributions of numerous independent random processes and can, therefore, be assumed to be Gaussian. The ML estimate of $d$ and $\theta$ for the case of exponential distribution was determined in \cite{jeske:ML}. Recently, the minimum variance unbiased estimate (MVUE) of the clock offset under an exponentially distributed network delay was proposed in \cite{qasim:mvue}. In this work, instead of working with a specific distribution, a general framework of the clock synchronization problem is proposed that yields a parameterized solution of the clock offset estimation problem in the classical as well Bayesian regime when the likelihood function of the observations, $U_j$ and $V_j$, is Gaussian, exponential or log-normally distributed.

In particular, the general notation used when the likelihood function of the observations $\mathbf{U}\overset{\Delta}{=}\left[U_1,\ldots,U_N\right]^T$ and $\mathbf{V}\overset{\Delta}{=}\left[V_1,\ldots,V_N\right]^T$ is Gaussian or log-normally distributed is given below.\\\\
Unconstrained Likelihood:
\begin{align}
f(\mathbf{U};\xi)&\propto \exp\left(\xi\sum_{j=1}^{N}\eta_{\xi}(U_j)-N\phi_{\xi}(\xi)\right)\label{U:uncons}\\
f(\mathbf{V};\psi)&\propto \exp\left(\psi\sum_{j=1}^{N}\eta_{\psi}(V_j)-N\phi_{\psi}(\psi)\right)\label{V:uncons}
\end{align}
where $\eta_{\xi}(U_j)$ and $\eta_{\psi}(V_j)$ are sufficient statistics for estimating $\xi$ and $\psi$, respectively. The log-partition functions $\phi_{\xi}(.)$ and $\phi_{\psi}(.)$ serve as normalization factors so that $f(\mathbf{U};\xi)$ and $f(\mathbf{V};\psi)$ are valid probability distributions. The likelihood function is called `unconstrained' since its domain is independent of the parameters $\xi$ and $\psi$.

Similarly, the general notation used for an exponentially distributed likelihood function is given below.\\\\
Constrained Likelihood:
\begin{align}
f(\mathbf{U};\xi)&\propto \exp\left(\xi\sum_{j=1}^{N}\eta_{\xi}(U_j)-N\phi_{\xi}(\xi)\right)\prod_{j=1}^{N}\mathbb{I}(U_j-\xi)\label{U:cons}\\
f(\mathbf{V};\psi)&\propto \exp\left(\psi\sum_{j=1}^{N}\eta_{\psi}(V_j)-N\phi_{\psi}(\psi)\right)\prod_{j=1}^{N}\mathbb{I}(V_j-\psi)
\label{V:cons}
\end{align}
where the indicator function $\mathbb{I}(.)$ is defined as
\begin{equation*}
\mathbb{I}(x )=\left\{\begin{matrix}
1 \quad x\geq0\\
0 \quad x<0
\end{matrix}\right.\;.
\end{equation*}
and the roles of $\eta_{\xi}(U_j)$, $\eta_{\psi}(V_j)$, $\phi_{\xi}(.)$ and $\phi_{\psi}(.)$ are similar to \eqref{U:uncons} and \eqref{V:uncons}. The likelihood function is called constrained since its domain depends on the parameters $\xi$ and $\psi$. It must be noted that the likelihood functions \eqref{U:uncons}-\eqref{V:cons} are expressed in terms of general exponential family distributions. This approach helps to keep the exposition sufficiently general and also allows us to recover the known results for the ML estimation of clock offset for Gaussian and exponentially distributed likelihood functions \cite{jeske:ML} \cite{eddie:novel}, and determine the ML estimator of the clock offset in case of log-normally distributed likelihood function, as shown in Section \ref{sec:ml}. The proposed approach will also prove useful in investigating a unified novel framework for clock offset estimation in the Bayesian setting for Gaussian, exponential or log-normally distributed likelihood functions, as will be shown in Section \ref{sec:factor}.

Some key ingredients of the proposed solution for the clock offset estimation problem, based on the properties of exponential family, can be summarized as follows \cite{wainwright:logpart}.
\begin{enumerate}
\item The mean and variance of the sufficient statistic $\eta_{\xi}(U_j)$ are expressed as
\begin{align}
\mathbb{E}\left[\eta_{\xi}(U_j)\right]&=\frac{\partial \phi_{\xi}(\xi)}{\partial \xi}\label{p:1}\\
\sigma^2_{\eta_{\xi}}\overset{\Delta}{=}\mathrm{Var}\left[\eta_{\xi}(U_j)\right]&=\frac{\partial^2 \phi_{\xi}(\xi)}{\partial \xi^2}\label{p:2}\;.
\end{align}
\item The moment generating function (MGF) of the statistic $\eta_{\xi}(U_j)$ is given by
\begin{equation}
M_{\eta_{\xi}}(h)=\exp\left(\phi_{\xi}(\xi+h)-\phi_{\xi}(\xi)\right)\label{mgf}\;.
\end{equation}
\item The non-negativity of the variance $\sigma^2_{\eta_{\xi}}$ in \eqref{p:2} implies that the log-partition function $\phi_{\xi}(.)$ is convex.
%\item The Taylor series expansion of the function $\phi_{\xi}(\xi)$ in neighborhood of 0 can be written as
%\begin{equation}
%\phi_{\xi}(\xi)=\phi(0)+\frac{\phi'(0)}{1!}(\xi-0)+\frac{\phi''(0)}{2!}(\xi-0)^2+\ldots\label{eq:13}
%\end{equation}
\item For Gaussian, exponential and log-normally distributed likelihood functions, the log-partition function $\phi_{\xi}(\xi)$ can be expressed as a second degree polynomial given by
    \begin{equation}
    \phi_{\xi}(\xi)= a_{\xi}\xi^2 \label{lpf:approx}\;.
    \end{equation}
     The coefficient $a_{\xi}$ in this approximation can be obtained using the variance of the statistic $\eta_{\xi}(U_j)$, which is assumed known. Using \eqref{p:2}, $a_\xi$ is given by
     \begin{equation*}
     a_{\xi}=\frac{\sigma^2_{\eta_{\xi}}}{2}\;.
     \end{equation*}
     If the statistical moment in \eqref{p:2} is not available, the empirical moment can be substituted since it readily follows from the weak law of large numbers that
\begin{align*}
\frac{\sum_{j=1}^{N}\eta_{\xi}(U_j)}{N}&\overset{p}{\rightarrow} \mathbb{E}\left[\eta_{\xi}(U)\right], \quad N \to \infty \\
\frac{\sum_{j=1}^{N}\eta^2_{\xi}(U_j)}{N}&\overset{p}{\rightarrow} \mathbb{E}\left[\eta^2_{\xi}(U)\right], \quad N \to \infty\;.
\end{align*}
\end{enumerate}
Similar expressions can also be written for $\eta_{\psi}(V_j)$, $M_{\eta_{\psi}}(h)$ and $\phi_{\psi}(\psi)$, respectively.
Using the aforementioned properties of the exponential family, the ML as well as Bayesian estimates of $\xi$ and $\psi$ are to be determined utilizing the data set $\{U_j,V_j\}_{j=1}^{N}$, based on \eqref{lpf:approx}.

\section{Maximum Likelihood Estimation}
In this section, the ML estimates\label{sec:ml} of $\theta$ are obtained by recasting the likelihood maximization as an instance of convex optimization. This approach differs from the graphical arguments used to maximize the likelihood in \cite{jeske:ML}. The specific cases of unconstrained and constrained likelihood functions are considered separately.
\subsection{Unconstrained Likelihood}
Using \eqref{U:uncons}, \eqref{V:uncons} and \eqref{lpf:approx}, the unconstrained likelihood functions are given by
\begin{align}
f(\mathbf{U};\xi)&\propto \exp\left(\xi\sum_{j=1}^{N}\eta_{\xi}(U_j)-N\frac{\sigma^2_{\eta_{\xi}}}{2}\xi^2\right)\label{U:uncons:approx}\\
f(\mathbf{V};\psi)&\propto \exp\left(\psi\sum_{j=1}^{N}\eta_{\psi}(V_j)-N\frac{\sigma^2_{\eta_{\psi}}}{2}\psi^2\right)\label{V:uncons:approx}\;.
\end{align}
The ML estimates of $\xi$ and $\psi$ can now be expressed as
\begin{align}
\hat\xi_{\textrm{ML}}&=\arg~\underset{\xi}{\max}\exp\left(\xi\sum_{j=1}^{N}\eta_{\xi}(U_j)-N
\frac{\sigma^2_{\eta_{\xi}}}{2}\xi^2\right) \label{xi:uncons}\\
\hat\psi_{\textrm{ML}}&= \arg~\underset{\psi}{\max}\exp\left(\psi\sum_{j=1}^{N}\eta_{\psi}(V_j)-N\frac{\sigma^2_{\eta_{\psi}}}{2}\psi^2\right)
\label{psi:uncons}\;.
\end{align}
\begin{theorem}
The likelihood\label{th:1} maximization problems \eqref{xi:uncons} and \eqref{psi:uncons} are strictly concave and the ML estimates are given by
\begin{subequations}
\begin{equation}
\hat\xi_{\textrm{ML}}=\frac{\sum_{j=1}^{N}\eta_{\xi}(U_j)}{N\sigma_{\eta_{\xi}}^2}\label{xi:psi:ML:a}
\end{equation}
\begin{equation}
\hat\psi_{\textrm{ML}}=\frac{\sum_{j=1}^{N}\eta_{\psi}(V_j)}{N\sigma_{\eta_{\psi}}^2} \label{xi:psi:ML:b}\;.
\end{equation}
\label{xi:psi:ML}
\end{subequations}
Hence, the ML estimator $\hat\theta_{\textrm{ML}}$ for the clock offset can be written as
\begin{equation}
\hat\theta_{\textrm{ML}}=\frac{\hat\xi_{\textrm{ML}}-\hat\psi_{\textrm{ML}}}{2}\label{theta:ML}\;.
\end{equation}
\end{theorem}
\begin{IEEEproof}
The ML estimate of $\xi$ in \eqref{xi:uncons} can be equivalently determined by maximizing the exponent in the likelihood function. It can be easily verified that the exponent in \eqref{xi:uncons}, which is a quadratic function of $\xi$, is strictly concave \cite{boyd:convex}. A similar explanation applies to the ML estimate of $\psi$. The ML estimates in \eqref{xi:psi:ML} can be obtained by setting the first derivative of the exponent with respect to $\xi$ (resp. $\psi$) to zero. The estimate $\hat\theta_{\textrm{ML}}$ in \eqref{theta:ML} can be obtained by invoking the invariance principle \cite{kay:est}. Hence, the proof readily follows.
\end{IEEEproof}

\subsubsection{Gaussian Distributed Likelihood Function}
A particular application of Theorem 1 is the case when the likelihood functions $f(U_j;\xi)$ and $f(V_j;\psi)$ have a Gaussian distribution i.e., $f(U_j;\xi)\sim\mathcal{N}(\xi,\sigma_{\xi}^2)$ and $f(V_j;\psi)\sim\mathcal{N}(\psi,\sigma_{\psi}^2)$ \cite{eddie:novel}. Hence, it follows that
\begin{equation}
f(\mathbf{U};\xi)=\frac{1}{(2\pi\sigma_\xi^2)^\frac{N}{2}}
\exp\left(-\frac{\sum_{j=1}^{N}(U_j-\xi)^2}{2\sigma_\xi^2}\right)\label{example:gauss:pdf}
\end{equation}
which can be rearranged as
\begin{equation*}
f(\mathbf{U};\xi)\propto \exp\left(\xi\frac{\sum_{j=1}^{N}U_j}{\sigma_\xi^2}-\frac{N}{2\sigma_\xi^2}\xi^2\right)
\;.\label{example:gauss:prop}
\end{equation*}
By comparing with \eqref{U:uncons:approx}, we have
\begin{equation}
\eta_{\xi}(U_j)=\frac{U_j}{\sigma_\xi^2}, \qquad \sigma_{\eta_{\xi}}^2=\frac{1}{\sigma_\xi^2}\label{example:gauss:param}
\end{equation}
and the ML estimate using \eqref{xi:psi:ML:a} is given by
\begin{equation}
\hat\xi_{\textrm{ML}}=\frac{\sum_{j=1}^{N}U_j}{N}\;.\label{example:gauss:xi:ML}
\end{equation}
By a similar reasoning, the ML estimate for $\psi$  \eqref{xi:psi:ML:b} is given by
\begin{equation*}
\hat\psi_{\textrm{ML}}=\frac{\sum_{j=1}^{N}V_j}{N}\;.
\end{equation*}
Using \eqref{theta:ML}, the ML estimate for the offset $\theta$ can be expressed as
\begin{equation}
\hat\theta_{\textrm{ML}}=\frac{\sum_{j=1}^{N}(U_j-V_j)}{2N}\;.\label{example:gauss:theta:ML}
\end{equation}
The above estimate coincides exactly with the one reported in \cite{eddie:novel}.

\subsubsection{Log-Normally Distributed Likelihood Function}
Another application of Theorem \ref{th:1} is when the samples $U_j$ and $V_j$ are log-normally distributed. In this case, we have
\begin{align}
f\left(\mathbf{U};\xi\right)&=\frac{1}{\sqrt{2\pi\sigma_\xi^2}}\prod_{j=1}^{N}U_j^{-1}
\exp\left(-\frac{\sum_{j=1}^{N}(\log U_j-\xi)}{2\sigma_\xi^2}\right)\nonumber \\
&\propto \exp\left(\xi\frac{\sum_{j=1}^{N}\log U_j}{\sigma_\xi^2}-\frac{N}{2\sigma_\xi^2}\xi^2\right)\;\label{example:lognormal:pdf}.
\end{align}
A comparison with \eqref{U:uncons:approx} yields
\begin{equation*}
\eta_{\xi}(U_j)=\frac{\log U_j}{\sigma_\xi^2}, \qquad \sigma_{\eta_{\xi}}^2=\frac{1}{\sigma_\xi^2}\;.\label{example:lognormal:param}
\end{equation*}
Following a similar line of reasoning,
\begin{equation*}
\eta_{\psi}(V_j)=\frac{\log V_j}{\sigma_\psi^2}, \qquad \sigma_{\eta_{\psi}}^2=\frac{1}{\sigma_\psi^2}\;.
\end{equation*}
The ML estimator for $\theta$ can obtained from \eqref{theta:ML} using \eqref{xi:psi:ML:a} and \eqref{xi:psi:ML:b}, and is given by
\begin{equation}
\hat\theta_{\textrm{ML}} = \frac{\sum_{j = 1}^{N} \left( \log U_j - \log V_j \right)}{N} \;.\label{example:lognormal:theta:ML}
\end{equation}

\subsection{Constrained Likelihood}
Using \eqref{U:cons}, \eqref{V:cons} and \eqref{lpf:approx}, the constrained likelihood functions are given by
\begin{align}
f(\mathbf{U};\xi)&\propto \exp\left(\xi\sum_{j=1}^{N}\eta_{\xi}(U_j)-N\frac{\sigma^2_{\eta_{\xi}}}{2}\xi^2\right)\prod_{j=1}^{N}
\mathbb{I}(U_j-\xi)
\label{U:cons:approx}\\
f(\mathbf{V};\psi)&\propto \exp\left(\psi\sum_{j=1}^{N}\eta_{\psi}(V_j)-N\frac{\sigma^2_{\eta_{\psi}}}{2}\psi^2\right)\prod_{j=1}^{N}
\mathbb{I}(V_j-\psi)\;.\label{V:cons:approx}
\end{align}
The resulting ML estimates of $\xi$ and $\psi$ can be obtained as
\begin{align}
\hat\xi_{\textrm{ML}}=\arg~\underset{\xi}{\max}&\exp\left(\xi\sum_{j=1}^{N}\eta_{\xi}(U_j)-N\frac{\sigma^2_{\eta_{\xi}}}
{2}\xi^2\right)\nonumber \\
\text{such that}\quad& U_j\geq\xi\label{xi:cons}\\
\hat\psi_{\textrm{ML}}= \arg~\underset{\psi}{\max}&\exp\left(\psi\sum_{j=1}^{N}\eta_{\psi}(V_j)-N\frac{\sigma^2_{\eta_{\psi}}}
{2}\psi^2\right)\nonumber \\
\text{such that}\quad&V_j\geq \psi\;.\label{psi:cons}
\end{align}
%for $j=1,\ldots,N$.
\begin{theorem}
The likelihood \label{th:2}maximization problems \eqref{xi:cons} and \eqref{psi:cons} are strictly concave and the ML estimates can be expressed as
\begin{subequations}
\begin{equation}
\hat\xi_{\textrm{ML}}=\min\left(\frac{\sum_{j=1}^{N}\eta_{\xi}(U_j)}{N\sigma_{\eta_{\xi}}^2},U_{(1)}\right)\label{xi:cons:ML}
\end{equation}
\begin{equation}
\hat\psi_{\textrm{ML}}=\min\left(\frac{\sum_{j=1}^{N}\eta_{\psi}(V_j)}{N\sigma_{\eta_{\psi}}^2}, V_{(1)}\right)\label{psi:cons:ML}
\end{equation}
\end{subequations}
where $U_{(1)}$ and $V_{(1)}$ denote the first order statistics of the samples $U_j$ and $V_j$, respectively. The ML estimator $\hat\theta_{\textrm{ML}}$ for the clock offset is given by
\begin{equation}
\hat\theta_{\textrm{ML}}=\frac{\hat\xi_{\textrm{ML}}-\hat\psi_{\textrm{ML}}}{2}\label{theta:cons:ML}\;.
\end{equation}
\end{theorem}
\begin{IEEEproof}
By using arguments similar to Theorem 1 and noting that the $N$ constrains $U_j\geq\xi$ (resp. $V_j\geq\psi$) are linear functions of $\xi$ (resp. $\psi$), the proof of concavity readily follows. Also, the likelihood maximization problems \eqref{xi:cons} and \eqref{psi:cons} can be equivalently expressed as
\begin{align*}
\hat\xi_{\textrm{ML}}=\arg~\underset{\xi}{\max}&\exp\left(\xi\sum_{j=1}^{N}\eta_{\xi}(U_j)-N\frac{\sigma^2_{\eta_{\xi}}}
{2}\xi^2\right)\nonumber \\
\text{such that}\quad& U_{(1)}\geq\xi\\
\hat\psi_{\textrm{ML}}=\arg~\underset{\psi}{\max}&\exp\left(\psi\sum_{j=1}^{N}\eta_{\psi}(V_j)-N
\frac{\sigma^2_{\eta_{\psi}}}{2}\psi^2\right)\nonumber \\
\text{such that}\quad&V_{(1)}\geq \psi\;.
\end{align*}
The unconstrained maximizer $\bar\xi$ of the objective function in \eqref{xi:cons} is given by \eqref{xi:psi:ML:a}. If $\bar\xi\leq U_{(1)}$, then $\hat\xi_{\textrm{ML}}=\bar\xi$. On the other hand, if $U_{(1)}\leq \bar\xi$, then the ML estimate is given by $\hat\xi_{\textrm{ML}}=U_{(1)}$ using concavity of the objective function. Combining the two cases, the ML estimate in \eqref{xi:cons:ML} is obtained. A similar explanation applies to the ML estimate $\hat\psi_{\textrm{ML}}$ in \eqref{psi:cons:ML} and $\hat\theta_{\textrm{ML}}$ follows from the invariance principle.
\end{IEEEproof}

\subsubsection{Exponentially Distributed Likelihood Function}
For the case when the likelihood functions are exponentially distributed, the density function of the samples $U_j$ can be written as \cite{jeske:ML}
\begin{equation}
f(\mathbf{U};\xi)=\lambda_\xi^N\exp\left(-\lambda_\xi\sum_{j=1}^{N}(U_j-\xi)\right)\mathbb{I}(U_{(1)}-\xi)
\label{example:exp:pdf}
\end{equation}
where $\lambda_\xi^{-1}$ is the mean of the delays $X_j$. The density function can be rearranged as
\begin{equation*}
f(\mathbf{U};\xi)\propto \exp\left(N\lambda_\xi\xi\right)\;.\label{example:exp:prop}
\end{equation*}
Comparing the above formulation with \eqref{U:cons:approx},
\begin{equation}
\eta_{\xi}(U_j)=\lambda_\xi, \qquad \sigma_\eta^2=0\;.\label{example:exp:param}
\end{equation}
Using \eqref{xi:cons:ML}, the ML estimate is given by
\begin{equation}
\hat\xi_{\textrm{ML}}=U_{(1)}\label{ml:E}\;.
\end{equation}
Employing a similar reasoning,
\begin{equation*}
\hat\psi_{\textrm{ML}}=V_{(1)}\;.
\end{equation*}
Using \eqref{theta:cons:ML}, the ML estimate of $\theta$ is given by
\begin{equation}
\hat\theta_{\textrm{ML}}=\frac{U_{(1)}-V_{(1)}}{2}\;,\label{example:exp:theta:ML}
\end{equation}
which coincides exactly with the one reported in \cite{jeske:ML}, where it is derived using graphical arguments.\\
\begin{remark}
The ML estimation method outlined above differs from the previous work \cite{jeske:ML} in that it is based on convex optimization, while \cite{jeske:ML} maximized the likelihood graphically. Hence, this approach presents an alternative view of the ML estimation of the clock offset. It also allows us the determine the ML estimator of $\theta$ when the likelihood function is log-normally distributed. In addition, Theorem \ref{th:2} will also be useful in Section \ref{sec:factor} where estimation of $\theta$ in the Bayesian regime is discussed .
\end{remark}

\section{A Factor Graph Approach}
The imperfections introduced by environmental conditions in the quartz oscillator in sensor nodes results in a time-varying clock offset between nodes in a WSN. To cater for such a temporal variation, in this section, \label{sec:factor}a Bayesian approach to the clock synchronization problem is adopted by representing the a-posteriori density as a factor graph. The inference is performed on the resulting factor graph by message passing using max-product algorithm. To ensure completeness, a brief description of factor graphs and the max-product algorithm is provided below.

A factor graph is a \emph{bipartite} graph that represents a factorization of a global function as a product of local functions called \emph{factors}, each factor being dependent on a subset of variables. Factor graphs are often used to produce a graphical model depicting the various inter-dependencies between a collection of interacting variables. Each factor is represented by a factor node and each variable has an edge or a half-edge. An edge connects a particular variable to a factor node if and only if it is an argument of the factor expressed by the factor node \cite{frank:factor}.

Inference can be performed by passing messages (sometimes called beliefs) along the edges of a factor graph. In particular, \emph{max-product} algorithm is used to compute the messages exchanged between variables and factor nodes. These messages can be summarized as follows\\
\newline$\mathbf{variable~to~factor~node:}$
\begin{equation}
m_{x\rightarrow f}\left ( x \right )=\prod _{h\in n\left ( x \right )\setminus {f}}m_{h\rightarrow x}\left ( x \right )\label{var:factor}
\end{equation}
$\mathbf{factor~node~to~variable:}$
\begin{equation}
m_{f\rightarrow x}\left ( x \right )=\max _{\setminus {\left \{ x \right \}}}\left(f\left ( Z \right )\prod _{z\in n\left ( f\right )\setminus {\left \{ x \right \}}}m_{z\rightarrow f}\left ( z \right )\right)\label{factor:var}
\end{equation}
where $Z=n\left(f\right)$ is the set of arguments of the local function $f$. The marginal distributions associated with each variable can be obtained by the product of all incoming messages on the variable.

\begin{figure*}[t!]
\begin{center}
\vspace{0.2in}
\includegraphics[scale=0.75]{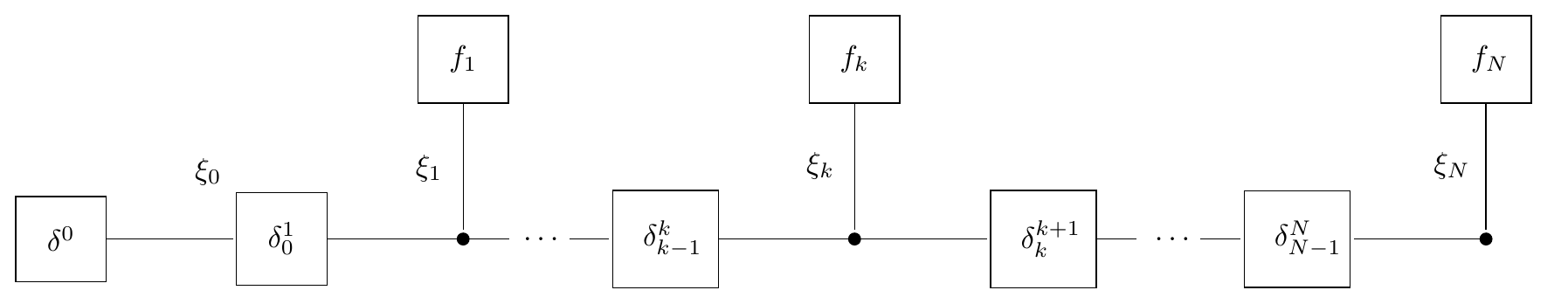}\vspace{0.35in}
\includegraphics[scale=0.75]{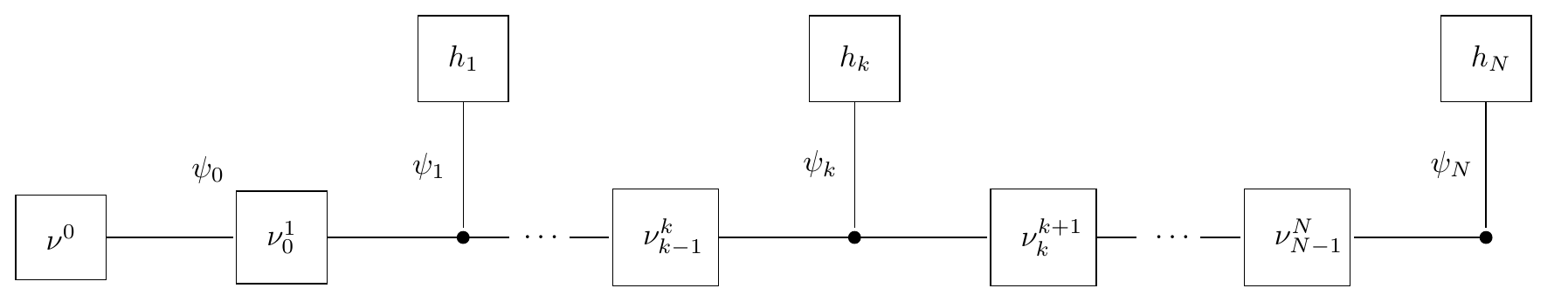}
\caption{Factor graph representation of the posterior density \eqref{pdf:pos}}
\label{fig:fact}
\end{center}
\end{figure*}

In order to sufficiently capture the temporal variations, the parameters $\xi$ and $\psi$ are assumed to evolve through a Gauss-Markov process given by
\begin{align*}
\xi_k&=\xi_{k-1}+w_k\\
\psi_k&=\psi_{k-1}+v_k \quad \text{for}~k=1,\ldots,N
\end{align*}
where $w_k$ and $v_k$ are $i.i.d$ such that $w_k,v_k\sim\mathcal{N}(0,\sigma^2)$. The posterior pdf can be expressed as
\begin{align}
f(\boldsymbol{\xi},\boldsymbol{\psi}|\boldsymbol{U},\boldsymbol{V})&\propto f(\boldsymbol{\xi},\boldsymbol{\psi})f(\boldsymbol{U},\boldsymbol{V}|\boldsymbol{\xi},\boldsymbol{\psi})\nonumber \\
&=f(\xi_0)\prod_{k=1}^{N}f(\xi_k|\xi_{k-1})f(\psi_0)\prod_{k=1}^{N}f(\psi_k|\psi_{k-1})\nonumber \\
&\cdot \prod_{k=1}^{N}f(U_k|\xi_k)f(V_k|\psi_k)\label{pdf:pos}
\end{align}
where uniform priors $f(\xi_0)$ and $f(\psi_0)$ are assumed. Define $\delta_{k-1}^{k}\overset{\Delta}{=}f(\xi_k|\xi_{k-1})\sim\mathcal{N}(\xi_{k-1},\sigma^2)$, $\nu_{k-1}^{k}\overset{\Delta}{=}f(\psi_k|\psi_{k-1})\sim\mathcal{N}(\psi_{k-1},\sigma^2)$, $f_k\overset{\Delta}{=}f(U_k|\xi_k)$, $h_k\overset{\Delta}{=}f(V_k|\psi_k)$, where the likelihood functions are given by
\begin{align}
f(U_k|\xi_k)&\propto\exp\left(\xi_k\eta_{\xi}(U_k)-\frac{\sigma^2_{\eta_{k}}}{2}\xi_k^2\right)\nonumber\\
f(V_k|\psi_k)&\propto\exp\left(\psi_k\eta_{\psi}(V_k)-\frac{\sigma^2_{\eta_{k}}}{2}\psi_k^2\right)
\label{pdf:uncons:bayes}
\;,
\end{align}
based on \eqref{lpf:approx}. The resulting factor graph representation of the posterior pdf is shown in Fig. \ref{fig:fact}.\\
\begin{remark}
A few important observations of this representation can be summarized below.
\begin{itemize}
\item Notice that the substitution in \eqref{subs} renders a cycle-free nature to the factor graph. Therefore, inference by message passing on such a factor graph is indeed optimal \cite{frank:factor}.
%\item Since we are only interested in estimating $\xi_N$ and $\psi_N$, it suffices to pass messages in one direction only.
\item The two factor graphs shown in Fig. 2 have a similar structure and hence, message computations will only be shown for the estimate $\hat\xi_N$. Clearly, similar expressions will apply to $\hat\psi_N$.
\end{itemize}
\end{remark}

In addition, only the case of constrained likelihood will be considered, since the case of an unconstrained likelihood is subsumed, as will be shown shortly. The clock offset estimator $\hat\theta_N$ can be obtained from $\hat\xi_N$ and $\hat\psi_N$ using \eqref{subs:theta}.

By defining $\alpha_{\xi,k}\overset{\Delta}{=}-\frac{\sigma^2_{\eta_{\xi,k}}}{2}$ and $\beta_{\xi,k}\overset{\Delta}{=}\eta_{\xi}(U_k)$, the constrained likelihood function for the samples $U_k$ can be written as
 \begin{equation}
f_k\propto \exp\left(\alpha_{\xi,k}\xi^2_k+\beta_{\xi,k}\xi_k\right)\mathbb{I}(U_k-\xi_k)\label{U:cons:bayes}\;.
\end{equation}
The message passing strategy starts by sending a message from the factor node $f_N$ to the variable $\xi_N$. The variable $\xi_N$ relays this message to the factor node $\delta_{N-1}^{N}$. The factor node computes the product of this message with the factor $\delta_{N-1}^{N}$ and sends the resulting message to the variable $\xi_{N-1}$ after `summarizing' over the variable $\xi_N$. In the max-product algorithm, a `max' function is used as a summary propagation operator (cf. \eqref{factor:var}). These messages are computed as
\begin{align*}
m_{f_N\rightarrow\xi_N}&=f_N\nonumber \\
m_{\xi_N\rightarrow\delta_{N-1}^{N}}&=f_N\nonumber& \\
m_{\delta_{N-1}^{N}\rightarrow\xi_{N-1}}&\propto \underset{\xi_N}{\max}~\delta_{N-1}^{N}\cdot m_{\xi_N\rightarrow\delta_{N-1}^{N}}\nonumber\\
&=\underset{\xi_N}{\max}~\frac{1}{\sqrt{2\pi\sigma^2}}\exp\left(\frac{-(\xi_N-\xi_{N-1})^2}{2\sigma^2}\right)
\nonumber \\
&\cdot \exp\left(\alpha_{\xi, N}\xi_N^2+\beta_{\xi,N}\xi_N\right) \mathbb{I}(U_N-\xi_N)\label{m:N}
\end{align*}
which can be rearranged as
\begin{equation}
\begin{split}
m_{\delta_{N-1}^{N}\rightarrow\xi_{N-1}}\propto \underset{\xi_N \leq U_N}{\max}~\exp &\big( A_{\xi,N}\xi_N^2+B_{\xi, N}\xi_{N-1}^2+ \\
&C_{\xi, N}\xi_N\xi_{N-1} +D_{\xi, N}\xi_N\big)
\end{split}
\label{m:delta:xi:N-1:max}
\end{equation}
where
\begin{align}
A_{\xi, N}&\overset{\Delta}{=}-\frac{1}{2\sigma^2}+\alpha_{\xi, N}, \quad B_{\xi, N}\overset{\Delta}{=}-\frac{1}{2\sigma^2}\nonumber \\
C_{\xi, N}&\overset{\Delta}{=}\frac{1}{\sigma^2}, \quad D_{\xi, N}\overset{\Delta}{=}\beta_{\xi,N} \label{const:N}\;.
\end{align}
Let $\bar\xi_N$ be the unconstrained maximizer of the exponent in the objective function above, i.e.,
\begin{equation}
\begin{split}
\bar\xi_N = \arg \underset{\xi_N}{\max}~\big( & A_{\xi, N}\xi_N^2+B_{\xi, N}\xi_{N-1}^2+C_{\xi, N}\xi_N\xi_{N-1}+  \\
& D_{\xi, N}\xi_N \big) \;.\nonumber \\
\end{split}
\end{equation}
This implies that
\begin{equation}
\bar\xi_N=-\frac{C_{\xi, N}\xi_{N-1}+D_{\xi, N}}{2A_{\xi, N}} \label{bar:N}\;.
\end{equation}
\begin{comment}
\begin{align}
\bar\xi_N = \arg \underset{\xi_N}{\max}~\big( & A_{\xi, N}\xi_N^2+B_{\xi, N}\xi_{N-1}^2+C_{\xi, N}\xi_N\xi_{N-1}+  \\
&+ D_{\xi, N}\xi_N \big) \nonumber \\
\Rightarrow~ \bar\xi_N=-&\frac{C_{\xi, N}\xi_{N-1}+D_{\xi, N}}{2A_{\xi, N}} \label{bar:N}\;.
\end{align}
\end{comment}
Following a line of reasoning similar to Theorem 2, it follows that
\begin{equation*}
\hat\xi_N=\min\left(\bar\xi_N,U_N\right)\label{hat:N}\;.
\end{equation*}
However, $\bar\xi_N$ depends on $\xi_{N-1}$, which is undetermined at this stage. Hence, we need to further traverse the chain backwards. Assuming that $\bar\xi_{N}\leq U_N$, $\bar\xi_N$ from \eqref{bar:N} can be plugged back in \eqref{m:delta:xi:N-1:max} which after some simplification yields
\begin{equation}
\begin{split}
m_{\delta_{N-1}^{N}\rightarrow\xi_{N-1}}\propto\exp\Bigg\{ &\left(B_{\xi, N}-\frac{C_{\xi, N}^2}{4A_{\xi, N}}\right)\xi^2_{N-1} - \\
& \frac{C_{\xi, N} D_{\xi, N}}{2A_{\xi, N}}\xi_{N-1} \Bigg\} \;.
\end{split}
\label{m:delta:xi:N-1}
\end{equation}
The message passed from the variable $\xi_{N-1}$ to the factor node $\delta_{N-2}^{N-1}$ is the product of the message \eqref{m:delta:xi:N-1} and the message received from the factor node $f_{N-1}$, i.e.,
\begin{equation*}
m_{\xi_{N-1}\rightarrow\delta_{N-2}^{N-1}}=m_{\delta_{N-1}^{N}\rightarrow\xi_{N-1}}\cdot m_{f_{N-1}\rightarrow\xi_{N-1}}\;.
\end{equation*}
Upon receipt of this message, the factor node $\delta_{N-2}^{N-1}$ delivers a product of this message and the factor $\delta_{N-2}^{N-1}$ to the variable node $\xi_{N-2}$ after summarizing over $\xi_{N-1}$. This message can be expressed as
\begin{align}
m&_{\delta_{N-2}^{N-1}\rightarrow\xi_{N-2}}\propto  \underset{\xi_{N-1} \leq U_{N-1}}{\max}~\delta_{N-2}^{N-1}\cdot m_{\xi_{N-1}\rightarrow\delta_{N-2}^{N-1}}\nonumber\\
=&\underset{\xi_{N-1}}{\max}~\frac{1}{\sqrt{2\pi\sigma^2}}\exp\left(-\frac{(\xi_{N-1}-\xi_{N-2})^2}{2\sigma^2}\right) \nonumber\\
& \cdot\exp\left \{ \left(B_{\xi, N}-\frac{C_{\xi, N}^2}{4A_{\xi, N}}\right)\xi^2_{N-1}-\frac{C_{\xi, N} D_{\xi, N}}{2A_{\xi, N}}\xi_{N-1} \right \}\nonumber\\
&\cdot\exp\left(\alpha_{\xi, N-1}\xi_{N-1}^2+\beta_{\xi,N-1}\xi_{N-1}\right)\mathbb{I}(U_{N-1}-\xi_{N-1})\nonumber \;.
\end{align}
After some algebraic steps, the message above can be compactly represented as
\begin{align}
m&_{\delta_{N-2}^{N-1} \rightarrow\xi_{N-2}} \propto\nonumber \underset{\xi_{N-1} \leq U_{N-1}} {\max}~\exp(A_{\xi, N-1}\xi_{N-1}^2+ \nonumber \\
&  B_{\xi, N-1}\xi_{N-2}^2+C_{\xi, N-1}\xi_{N-1}\xi_{N-2}+D_{\xi, N-1}\xi_{N-1}) \label{m:delta:xi:N-2:max}
\end{align}
where
\begin{align*}
A_{\xi, N-1}&\overset{\Delta}{=}-\frac{1}{2\sigma^2}+\alpha_{\xi, N-1}+B_{\xi, N}-\frac{C_{\xi, N}^2}{4A_{\xi, N}}, \quad \nonumber \\
B_{\xi, N-1}&\overset{\Delta}{=}-\frac{1}{2\sigma^2}, \quad C_{\xi, N-1}\overset{\Delta}{=}\frac{1}{\sigma^2} \\
D_{\xi, N-1}&\overset{\Delta}{=}\beta_{\xi,N-1}-\frac{C_{\xi, N}D_{\xi, N}}{2A_{\xi, N}}\;.\label{const:N-1}
\end{align*}
Proceeding as before, the unconstrained maximizer $\bar\xi_{N-1}$ of the objective function above is given by
\begin{equation*}
\bar\xi_{N-1}=-\frac{C_{\xi, N-1}\xi_{N-2}+D_{\xi, N-1}}{2A_{\xi, N-1}}\label{bar:N-1}
\end{equation*}
and the solution to the maximization problem \eqref{m:delta:xi:N-2:max} is expressed as
\begin{equation*}
\hat\xi_{N-1}=\min\left(\bar\xi_{N-1},U_{N-1}\right)\;.\label{hat:N-1}
\end{equation*}
Again, $\bar\xi_{N-1}$ depends on $\xi_{N-2}$ and therefore, the solution demands another traversal backwards on the factor graph representation in Fig. 2. By plugging $\bar\xi_{N-1}$ back in \eqref{m:delta:xi:N-2:max}, it follows that
\begin{align}
&m_{\delta_{N-2}^{N-1}\rightarrow\xi_{N-2}}\propto\nonumber \\
&\exp\left \{ \left(B_{\xi, N-1}-\frac{C_{\xi, N-1}^2}{4A_{\xi, N-1}}\right)\xi^2_{N-2}-\frac{C_{\xi, N-1} D_{\xi, N-1}}{2A_{\xi, N-1}}\xi_{N-2} \right \}\label{m:delta:xi:N-2}
\end{align}
which has a form similar to \eqref{m:delta:xi:N-1}. It is clear that one can keep traversing back in the graph yielding messages similar to \eqref{m:delta:xi:N-1} and \eqref{m:delta:xi:N-2}. In general, for $i = 1, \ldots, N-1$ we can write
\begin{equation}
\begin{split}
A_{\xi, N-i} &\overset{\Delta}{=} -\frac{1}{2 \sigma^2} + \alpha_{\xi, N-i} + B_{\xi, N-i+1} - \frac{C_{\xi, N-i+1}^2}{4 A_{\xi, N-i+1}} \\
B_{\xi, N-i} & \overset{\Delta}{=} -\frac{1}{2 \sigma^2}, \quad C_{\xi, N-i} \overset{\Delta}{=} \frac{1}{\sigma^2} \\
D_{\xi, N - i} & \overset{\Delta}{=} \beta_{\xi,N - i} - \frac{C_{\xi, N - i + 1} D_{\xi, N - i + 1}}{2 A_{\xi, N - i + 1}}
\end{split}
\label{const:N-i}
\end{equation}
and
\begin{eqnarray}
\bar{\xi}_{N-i} &=& -\frac{C_{\xi, N-i} \xi_{N - i - 1} + D_{\xi, N - i}}{2 A_{\xi, N - i}} \label{bar:N-i} \\
\hat{\xi}_{N-i} &=& \min \left( \bar{\xi}_{N-i}, U_{N - i} \right) \label{hat:N-i}\;.
\end{eqnarray}
Using \eqref{bar:N-i} and \eqref{hat:N-i} with $i = N-1$, it follows that
\begin{align}
\bar\xi_{1}=-\frac{C_{\xi, 1}\xi_{0}+D_{\xi, 1}}{2A_{\xi, 1}} \label{bar:1}\\
\hat\xi_{1}=\min\left(\bar\xi_{1},U_{1}\right)\label{hat:1}\;.
\end{align}
Similarly, by observing the form of \eqref{m:delta:xi:N-1} and \eqref{m:delta:xi:N-2}, it follows that
\begin{equation}
m_{\delta_{0}^{1}\rightarrow\xi_{0}}\propto \exp\left \{ \left(B_{\xi, 1}-\frac{C_{\xi,1}^2}{4A_{\xi,1}}\right)\xi^2_{0}-\frac{C_{\xi,1} D_{\xi,1}}{2A_{\xi,1}}\xi_{0} \right \}\;.\label{m:delta:xi:0}
\end{equation}
The estimate $\hat\xi_0$ can be obtained by maximizing the received message in \eqref{m:delta:xi:0}. It can be noticed from the structure of the factor graph that this maximization is inherently unconstrained i.e.,
\begin{align}
\hat\xi_0=\bar\xi_0&=\underset{\xi_{0}}{\max}~m_{\delta_{0}^{1}\rightarrow\xi_{0}}\nonumber\\
\Rightarrow \hat\xi_0&=\frac{C_{\xi,1}D_{\xi, 1}}{4A_{\xi, 1}B_{\xi, 1}-C_{\xi, 1}^2}\;.\label{hat:0}
\end{align}
The estimate in \eqref{hat:0} can now be substituted in \eqref{bar:1} to yield $\bar\xi_1$, which can then be used to solve for $\hat\xi_1$ in \eqref{hat:1}. Clearly, this chain of calculations can be continued using recursions \eqref{bar:N-i} and \eqref{hat:N-i}.\\
\begin{comment}
Clearly, the unconstrained maximizer $\bar\xi_2$ for the state $\xi_2$ is given by
\begin{equation}
\bar\xi_{2}=-\frac{C_{\xi, 2}\xi_{1}+D_{\xi, 2}}{2A_{\xi, 2}}\label{bar:2}
\end{equation}
which can be obtained by using the estimate $\hat\xi_1$ from \eqref{hat:1}. The estimate $\hat\xi_2$ can then be expressed using $\bar\xi_2$ as
\begin{equation}
\hat\xi_2=\min\left(\bar\xi_2,U_2\right)
\end{equation}
\end{comment}
Define
\begin{equation}
g_{\xi,k}(x)\overset{\Delta}{=}-\frac{C_{\xi,k}x+D_{\xi,k}}{2A_{\xi,k}}\;.\label{g:def}
\end{equation}
A key property of the function $g_{\xi,k}(.)$, which proves useful in the quest for a closed form solution, can be summarized in the following lemma.
\begin{lemma}
For real \label{lem:1}numbers $a$ and $b$, the function $g_{\xi,k}(.)$ defined in \eqref{g:def} satisfies
\begin{equation*}
g_{\xi,k}\left(\min(a, b)\right)=\min\left(g_{\xi,k}(a), g_{\xi,k}(b)\right)\;.
\end{equation*}
\end{lemma}
\begin{IEEEproof}
The constants $A_{\xi,k}$, $C_{\xi,k}$ and $D_{\xi,k}$ are defined in \eqref{const:N} and \eqref{const:N-i}. The proof follows by noting that $\frac{-C_{\xi,k}}{2A_{\xi,k}}>0$ which implies that $g_{\xi,k}(.)$ is a monotonically increasing function.
\end{IEEEproof}
With the notation $g_{\xi,k}(.)$, the following chain of equalities can be conveniently written as
\begin{align*}
\bar\xi_1&=g_{\xi,1}\left(\hat\xi_0\right)\nonumber \\
\hat\xi_1&=\min\left(U_1,g_{\xi,1}\left(\hat\xi_0\right)\right)\nonumber \\
\bar\xi_2&=g_{\xi,2}\left(\hat\xi_1\right)\nonumber \\
\hat\xi_2&=\min\left(U_2,g_{\xi,2}\left(\hat\xi_1\right)\right)
\end{align*}
where
\begin{align}
g_{\xi,2}\left(\hat\xi_1\right)&=g_{\xi,2}\left(\min\left(U_1,g_{\xi,1}\left(\hat\xi_0\right)\right)\right)\nonumber \\
&=\min\left(g_{\xi,2}\left(U_1\right),g_{\xi,2}\left(g_{\xi,1}\left(\hat\xi_0\right)\right)\right)\label{g:nest}
\end{align}
where \eqref{g:nest} follows from Lemma \ref{lem:1}. The estimate $\hat\xi_2$ can be expressed as
\begin{align*}
\hat\xi_2&=\min\left(U_2,\min\left(g_{\xi,2}\left(U_1\right),g_{\xi,2}\left(g_{\xi,1}\left(\hat\xi_0\right)\right)\right)\right)
\nonumber\\
&=\min\left(U_2,g_{\xi,2}\left(U_1\right),g_{\xi,2}\left(g_{\xi,1}\left(\hat\xi_0\right)\right)\right)\;.
\end{align*}
By following the same procedure, one can write
\begin{equation*}
\begin{split}
\hat\xi_3=\min\Big(& U_3,g_{\xi,3}\left(U_2\right),g_{\xi,3}\left(g_{\xi,2}\left(U_1\right)\right), \\
& g_{\xi,3}\left(g_{\xi,2}\left(g_{\xi,1}\left(\hat\xi_0\right)\right)\right)\Big)\;.
\end{split}
\label{hat:3:g}
\end{equation*}
For $m \geq j$, define
\begin{equation}
G_{\xi, j}^m(.)\overset{\Delta}{=}g_{\xi,m}\left(g_{\xi,m-1}\ldots g_{\xi,j}\left(.\right)\right)\;.\label{G:def}
\end{equation}
The estimate $\hat\xi_3$ can, therefore, be compactly represented as
\begin{equation*}
\hat\xi_3=\min\left(U_3, G_{\xi,3}^3\left(U_2\right),G_{\xi,2}^3\left(U_1\right), G_{\xi,1}^3\left(\hat\xi_0\right)\right)\;.
\end{equation*}
Hence, one can keep estimating $\hat\xi_k$ at each stage using this strategy. Note that the estimator only depends on functions of data and can be readily evaluated.

In order to derive analogous expressions for $\psi$, a similar line of reasoning should be followed. In particular, the constants $A_{\xi, N-i}$, $B_{\xi, N-i}$, $C_{\xi, N-i}$ and $D_{\xi, N-i}$ for $i = 0, \ldots, N-1$, can be obtained straightforwardly from \eqref{const:N} and \eqref{const:N-i} by substituting $\alpha_{\xi, N-i}$ and $\beta_{\xi, N-i}$ with $\alpha_{\psi, N-i}$ and $\beta_{\psi, N-i}$, respectively. Using these constants, $\hat\psi_0$, $g_{\psi, k}$ and $G_{\psi, j}^m$ can be defined analogously to \eqref{hat:0}, \eqref{g:def} and \eqref{G:def}.

Generalizing this framework, the closed form expression for the clock offset estimate $\hat\theta_N$ is given by the following theorem.\\
\begin{theorem}
The state estimates\label{th:3} $\hat\xi_N$ and $\hat\psi_N$ for the posterior pdf in \eqref{pdf:pos} can be expressed as
%\begin{subequations*}
\begin{align*}
\hat\xi_N&=\min\left(U_N,G_{\xi,N}^N\left(U_{N-1}\right),\dots,G_{\xi,2}^N\left(U_1\right),
G_{\xi,1}^N\left(\hat\xi_0\right)\right)\\
%\end{equation*}
%\begin{equation*}
\hat\psi_N&=\min\left(V_N,G_{\psi,N}^N\left(V_{N-1}\right),\dots,G_{\xi,2}^N\left(V_1\right),
G_{\xi,1}^N\left(\hat\psi_0\right)\right)
\end{align*}
%\end{subequations*}
and the factor graph based clock offset estimate (FGE) $\hat\theta_N$ is given by
\begin{equation}
\hat\theta_N=\frac{\hat\xi_N-\hat\psi_N}{2}\;.\label{theta:fg}
\end{equation}
\end{theorem}
\begin{IEEEproof}
The proof follows from the discussion above and using \eqref{subs:theta}.
\end{IEEEproof}\vspace{0.15in}
\begin{remark}
The closed form expressions in Theorem \ref{th:3} enable the estimation of the clock offset, when it may be time-varying and the likelihood functions, $f(U_k|\xi_k)$ and $f(V_k|\psi_k)$, have a Gaussian, exponential or log-normal distribution.
\end{remark}
\subsection{Gaussian Distributed Likelihood Function}
A particular \label{exampleb:gauss}case of the Bayesian framework described above occurs when likelihood functions $f(U_k|\xi_k)$ and $f(V_k|\psi_k)$ have a Gaussian distribution, i.e., $f(U_k|\xi_k)\sim\mathcal{N}(\xi_k,\sigma_{\xi,k}^2)$ and $f(V_k|\psi_k)\sim\mathcal{N}(\psi_k,\sigma_{\psi,k}^2)$, i.e.,
\begin{align}
f\left ( U_k|\xi_k \right )&=\frac{1}{\sqrt{2\pi\sigma_{\xi,k}^2}}\exp\left \{ -\frac{\left ( U_k-\xi_k \right )^2}{2\sigma_{\xi,k}^2} \right \}\nonumber \\
&\propto \exp\left(\frac{\xi_k U_k}{2\sigma_{\xi,k}^2}-\frac{\xi_k^2}{2\sigma_{\xi,k}^2}\right)\;.\label{pdf:gauss:k}
\end{align}
The aforementioned Gaussian distribution constitutes an unconstrained likelihood function, i.e., the domain of the pdf is independent of the unknown parameter $\xi_k$. It is clear from the message passing approach that at each stage $k$ of the factor graph, the unconstrained maximizer $\bar\xi_k$ is the actual solution to the likelihood maximization problem
\begin{equation*}
\underset{\xi_k}{\max}~\exp\left(A_{\xi,k}\xi_k^2+B_{\xi,k}\xi_{k-1}^2+C_{\xi,k}\xi_k\xi_{k-1}+D_{\xi,k}\xi_k\right)
\end{equation*}
i.e., $\hat\xi_k=\bar\xi_k~\forall k=1,\ldots,N$. Hence, the unconstrained likelihood maximization problem is subsumed in the message passing framework for constrained likelihood maximization. It follows from Theorem \ref{th:3} that $\hat\xi_N$ for Gaussian distributed observations $U_k$ in \eqref{pdf:gauss:k} is given by
\begin{equation*}
\hat\xi_N=G_{\xi,1}^N\left(\hat\xi_0\right)
\end{equation*}
where $\hat\xi_0$ and $G_{\xi,1}^N(.)$ are defined in \eqref{hat:0} and \eqref{G:def}, respectively. Evaluating $\hat\xi_N$ requires to determine the constants in \eqref{const:N} and \eqref{const:N-i}. By comparing \eqref{pdf:gauss:k} with \eqref{U:cons:bayes}, we have
\begin{equation}
\alpha_{\xi,k}=-\frac{1}{2\sigma_{\xi,k}^2}, \quad \beta_{\xi,k}=\frac{U_k}{\sigma_{\xi,k}^2}\;.\label{gauss:param:k}
\end{equation}
Using these values for $\alpha_{\xi,k}$ and $\beta_{\xi,k}$, \eqref{const:N} and \eqref{const:N-i} can be written as
\begin{align}
A_{\xi,N}&=-\frac{1}{2\sigma^2}-\frac{1}{2\sigma_{\xi, N}^2}, \quad B_{\xi,N}=-\frac{1}{2\sigma^2}\nonumber \\
C_{\xi,N}&=\frac{1}{\sigma^2}, \quad D_{\xi,N}=\frac{U_N}{\sigma_{\xi, N}^2}\label{const:G:N} \\
A_{\xi, N-i}&=-\frac{1}{2\sigma^2}-\frac{1}{2\sigma_{\xi, N-i}^2}+B_{\xi, N-i+1}-\frac{C_{\xi, N-i+1}^2}{4A_{\xi, N-i+1}}\nonumber \\ B_{\xi, N-i}&=-\frac{1}{2\sigma^2},  \quad C_{\xi, N-i}=\frac{1}{\sigma^2}\nonumber \\
D_{\xi, N-i}&=\frac{U_{N-i}}{\sigma_{\xi, N-i}^2}-\frac{C_{\xi, N-i+1}D_{\xi, N-i+1}}{2A_{\xi, N-i+1}}\nonumber
\end{align}
for $i=1, \ldots,N-1$. Using similar arguments, it can be shown that the estimate $\hat\psi_N$ is given by
\begin{equation*}
\hat\psi_N=G_{\psi,1}^N\left(\hat\psi_0\right) \;.
\end{equation*}
It follows from \eqref{subs:theta} that the FGE, $\hat\theta_N$, can be expressed as
\begin{equation}
\hat\theta_N= \frac{G_{\xi,1}^N\left(\hat\xi_0\right) - G_{\psi,1}^N\left(\hat\psi_0\right)}{2} \label{FGE_theta_gauss}\;.
\end{equation}
 %It must be noted that the estimator $\hat\xi_N$ so evaluated, coincides with the MAP estimator since the log-partition function for a Gaussian distribution is a second degree polynomial in $\xi$, as assumed in \eqref{lpf:approx}. Similar results apply to the estimator $\hat\psi_N$.

The behavior of $\hat\theta_N$ can be further investigated for the case when the noise variance $\sigma^2$ in the Gauss-Markov model goes to zero. Consider
\begin{equation*}
g_{\xi,N}(\xi)=-\frac{C_{\xi,N}\xi+D_{\xi,N}}{2A_{\xi,N}}
\end{equation*}
where the constants $A_{\xi,N}$, $B_{\xi,N}$, $C_{\xi,N}$ and $D_{\xi,N}$ are given by \eqref{const:G:N}. After some algebraic steps, we have
\begin{equation*}
g_{\xi,N}(\xi)=\frac{\sigma_\xi^2\xi+\sigma^2U_N}{\sigma_\xi^2+\sigma^2} \;.
\end{equation*}
As $\sigma^2\rightarrow 0$, $g_{\xi,N}(\xi)\rightarrow \xi$. Similarly, it can be shown that $g_{\xi, N-1}(\xi)\rightarrow \xi$ as $\sigma^2\rightarrow 0$. Hence, it follows that in the low system noise regime, as $\sigma^2\rightarrow 0$
\begin{equation*}
\hat\xi_N\rightarrow \hat\xi_0=\frac{C_{\xi,1}D_{\xi,1}}{4A_{\xi,1}B_{\xi,1}-C_{\xi,1}^2}\;.
\end{equation*}
Similarly, it can be shown that
\begin{equation*}
\hat\psi_N\rightarrow \hat\psi_0=\frac{C_{\psi,1}D_{\psi,1}}{4A_{\psi,1}B_{\psi,1}-C_{\psi,1}^2}\;.
\end{equation*}
Therefore
\begin{equation*}
\hat\theta_N\rightarrow \frac{\hat\xi_0 - \hat\psi_0}{2}\;,
\end{equation*}
which can be proven to be equal to the ML estimator \eqref{example:gauss:theta:ML}.

\subsection{Log-Normally Distributed Likelihood Function}
The log-normally distributed likelihood function in the Bayesian regime can be expressed as
\begin{align}
f(U_k | \xi_k) &= \frac{1}{U_k \sigma_{\xi,k} \sqrt{2 \pi}} \exp \left( -\frac{\left( \log U_k - \xi_k \right)^2}{2 \sigma_{\xi,k}^2} \right)\nonumber \\
&\propto \exp\left(\frac{\xi_k \log(U_k)}{2\sigma_{\xi,k}^2}-\frac{\xi_k^2}{2\sigma_{\xi,k}^2}\right) \;.\label{exampleb:lognormal:pdf}
\end{align}
By comparing \eqref{exampleb:lognormal:pdf} and \eqref{U:cons:bayes}, we have
\begin{equation*}
\alpha_{\xi,k} = -\frac{1}{2 \sigma_{\xi,k}^2} \quad \beta_{\xi,k} = \frac{\log U_k}{\sigma_{\xi,k}^2}\;.
\end{equation*}
Clearly, the only difference here with the Gaussian distribution is a redefinition of $\beta_{\xi,k}$. The expression of $\hat\xi_N$ in this case is again
\begin{equation*}
\hat\xi_N=G_{\xi,1}^N\left(\hat\xi_0\right)
\end{equation*}
where $G_{\xi,1}^N(.)$ and $\hat\xi_0$ are given by \eqref{G:def} and \eqref{hat:0}, respectively.
 %Assuming the log-normally distributed likelihood \eqref{exampleb:lognormal:pdf},
The recursively evaluated constants in \eqref{const:N} and \eqref{const:N-i} can be written as
\begin{align*}
A_{\xi,N}&=-\frac{1}{2\sigma^2}-\frac{1}{2\sigma_{\xi, N}^2}, \quad B_{\xi,N}=-\frac{1}{2\sigma^2}\nonumber \\
C_{\xi,N}&=\frac{1}{\sigma^2}, \quad D_{\xi,N}=\frac{\log U_N}{\sigma_{\xi, N}^2} \\
A_{\xi, N-i}&=-\frac{1}{2\sigma^2}-\frac{1}{2\sigma_{\xi, N-i}^2}+B_{\xi, N-i+1}-\frac{C_{\xi, N-i+1}^2}{4A_{\xi, N-i+1}}\nonumber \\ B_{\xi, N-i}&=-\frac{1}{2\sigma^2},  \quad C_{\xi, N-i}=\frac{1}{\sigma^2}\nonumber \\
D_{\xi, N-i}&=\frac{\log U_{N-i}}{\sigma_{\xi, N-i}^2}-\frac{C_{\xi, N-i+1}D_{\xi, N-i+1}}{2A_{\xi, N-i+1}}
\end{align*}
for $i=1, \ldots,N-1$. Similar arguments apply to $\psi$. Hence, the FGE $\hat\theta_N$ can be expressed as
\begin{equation}
\hat\theta_N= \frac{G_{\xi,1}^N\left(\hat\xi_0\right) - G_{\psi,1}^N\left(\hat\psi_0\right)}{2} \label{FGE_theta_logn}\;.
\end{equation}
Again, as the Gauss-Markov system noise $\sigma^2\rightarrow 0$, the above estimator approaches its ML counterpart \eqref{example:lognormal:theta:ML}.

\subsection{Exponential Distribution}
Theorem \ref{th:3} can also be used to derive a Bayesian estimator $\hat\xi_N$ for the exponentially distributed likelihood case considered in \cite{jeske:ML}. In this case, we have
\begin{align}
f(U_k|\xi_k)&=\lambda_{\xi}\exp\left(-\lambda_{\xi}(U_k-\xi_k)\right)\mathbb{I}(U_k-\xi_k)\nonumber \\
&\propto \exp(\lambda_{\xi}\xi_k)\mathbb{I}(U_k-\xi_k)\label{U:exp}
\end{align}
where $\lambda_{\xi}^{-1}$ is the mean network delay of $X_k$. A comparison of \eqref{U:exp} with \eqref{U:cons:bayes} reveals that
\begin{equation*}
\alpha_{\xi,k}=0, \quad \beta_{\xi,k}=\lambda_{\xi}\;.
\end{equation*}
For these values of $\alpha_{\xi,k}$ and $\beta_{\xi,k}$, the constants $A_{\xi,k}$, $B_{\xi,k}$, $C_{\xi,k}$ and $D_{\xi,k}$ are given by
\begin{align*}
A_{\xi,k}&=-\frac{1}{2\sigma^2}, \quad B_{\xi,k}=-\frac{1}{2\sigma^2}\nonumber\\
C_{\xi,k}&=\frac{1}{\sigma^2}, \quad D_{\xi,k}=\lambda_{\xi} \label{const:E:i}
\end{align*}
for all $k=1,\ldots,N$. Using Theorem \ref{th:3}, we have
\begin{align*}
G_{\xi,N}^N(U_{N-1})&=-\frac{C_{\xi,N}U_{N-1}+D_{\xi,N}}{2A_{\xi,N}}\\
&=U_{N-1}+\lambda_{\xi}\sigma^2\;.
\end{align*}
Similarly it can be shown that
\begin{equation*}
G_{\xi,N-1}^N(U_{N-2})=U_{N-2}+2\lambda_{\xi}\sigma^2
\end{equation*}
and so on. The estimator $\hat\xi_0$ at the last step can be evaluated as
\begin{equation*}
\hat\xi_0=\frac{C_{\xi,1}D_{\xi,1}}{4A_{\xi,1}B_{\xi,1}-C_{\xi,1}^2}=+\infty\;.
\end{equation*}
This implies that
\begin{equation}
G_{\xi,1}^N(\hat\xi_0)=+\infty\;. \label{G:infty}
\end{equation}
Using \eqref{G:infty} and Theorem \ref{th:3}, it readily follows that
\begin{align}
\hat\xi_N=\min(U_N, U_{N-1}&+\lambda_{\xi}\sigma^2,U_{N-2}+2\lambda_{\xi}\sigma^2,\nonumber \\
&\ldots,U_1+(N-1)\lambda_{\xi}\sigma^2)\;.\label{map:E}
\end{align}
Similarly, for the pdf \cite{jeske:ML}
\begin{align*}
f(V_k|\psi_k)&=\lambda_{\psi}\exp\left(-\lambda_{\psi}(V_k-\psi_k)\right)\mathbb{I}(V_k-\psi_k)\nonumber \\
&\propto \exp(\lambda_{\psi}\psi_k)\mathbb{I}(V_k-\psi_k)\;,
\end{align*}
the estimate $\hat\psi_N$ is given by
\begin{align}
\hat\psi_N=\min(V_N, V_{N-1}&+\lambda_{\psi}\sigma^2,V_{N-2}+2\lambda_{\psi}\sigma^2,\nonumber \\
&\ldots,V_1+(N-1)\lambda_{\psi}\sigma^2)\;\label{map:E:psi}
\end{align}
and the estimate $\hat\theta_N$ can be obtained using \eqref{theta:fg}, \eqref{map:E} and \eqref{map:E:psi} as
\begin{align}
\hat\theta_N=\frac{1}{2} \min(U_N, U_{N-1}&+\lambda_{\xi}\sigma^2,U_{N-2}+2\lambda_{\xi}\sigma^2,\nonumber \\
&\ldots,U_1+(N-1)\lambda_{\xi}\sigma^2)-\nonumber\\\frac{1}{2}\min(V_N, V_{N-1}&+\lambda_{\psi}\sigma^2,V_{N-2}+2\lambda_{\psi}\sigma^2,\nonumber \\
&\ldots,V_1+(N-1)\lambda_{\psi}\sigma^2)\;.\label{example:exp:theta:N}
\end{align}
%\begin{equ
As the Gauss-Markov system noise $\sigma^2 \rightarrow 0$, \eqref{example:exp:theta:N} yields
\begin{equation*}
\hat\theta_N\rightarrow \hat\theta_{\textrm{ML}}=\frac{\min\left(U_N, \ldots, U_1\right)-\min\left(V_N, \ldots, V_1\right)}{2}\;,
\end{equation*}
which is the ML estimator given by \eqref{example:exp:theta:ML}.

\section{Classical and Bayesian Bounds}
To evaluate the performance \label{sec:bounds}of the estimators derived in the preceding sections, classical as well as Bayesian lower bounds on the variance of the estimators are discussed. The placement of a lower bound allows one to compare estimators by plotting their performance against the bound. It must be emphasized here that the results in this section assume no specific form of the log-partition function and are therefore, valid for arbitrary distributions from the exponential family, which is a wide class and contains almost all distributions of interest. Hence, these results are fairly general and can be useful in their own right in classical as well as Bayesian parameter estimation theory, and at the same time will be used as a stepping stone towards comparing the estimators developed thus far.

The likelihood function of the data is considered an arbitrary member of the exponential family of distributions. In addition, depending on whether the domain of the likelihood depends on the parameter to be estimated, both cases of unconstrained as well as constrained likelihood functions are discussed to maintain full generality. The general expressions for the unconstrained and constrained likelihood functions for observations $\mathbf{Z}\overset{\Delta}{=}\left[Z_1,\ldots,Z_N\right]^T$ are given by \\\\
Unconstrained Likelihood:
\begin{equation}
f(\mathbf{Z};\rho)\propto \exp\left(\rho\sum_{j=1}^{N}\eta(Z_j)-N\phi(\rho)\right)\label{U:uncons:recall}
\end{equation}
Constrained Likelihood:
\begin{equation}
f(\mathbf{Z};\rho)\propto \exp\left(\rho\sum_{j=1}^{N}\eta(Z_j)-N\phi(\rho)\right)\prod_{j=1}^{N}\mathbb{I}(Z_j-\rho)
\label{U:cons:recall}
\end{equation}
where $\rho$ is the scalar parameter to be estimated. The goal is to derive lower bounds on the variance of estimators of $\rho$. For the case of classical estimation, the Cramer-Rao and Chapman-Robbins bounds are considered, while the Bayesian Cramer-Rao bound and a Bayesian version of the Chapman-Robbins bound are derived for the Bayesian paradigm.
\subsection{Cramer-Rao Lower Bound}
The Cramer-Rao lower bound (CRB) is a lower bound on the variance of an unbiased estimator of a deterministic parameter. It is useful primarily because it is relatively simple to compute. However, it relies on certain `regularity conditions' which are not satisfied by constrained likelihood functions when the domain of the likelihood depends on the unknown parameter (cf. \eqref{U:cons:recall}). Hence, CRB is determined for the case of unconstrained likelihood functions only.

In particular, CRB states that the variance of an unbiased estimator of $\rho$ is lower bounded by
\begin{equation}
\mathrm{Var}(\hat\rho)\geq \frac{-1}{\mathbb{E}\left[\frac{\partial^2 \ln f\left ( \mathbf{Z};\rho \right ) }{\partial \rho^2}\right]}\;.\label{crlb}
\end{equation}
\begin{theorem}
The CRB for $\rho$ in the unconstrained likelihood function in \eqref{U:uncons:recall} is given by
\begin{equation}
\mathrm{Var}(\hat\rho)\geq \frac{1}{N\sigma_{\eta}^2}\label{crlb:our}
\end{equation}
where
\begin{equation*}
\sigma_{\eta}^2=\frac{\partial^2 \phi\left(\rho \right ) }{\partial \rho^2}\;.
\end{equation*}
\end{theorem}
\begin{IEEEproof}
The Fisher information for the likelihood function is given by
\begin{align*}
I(\rho)&\overset{\Delta}{=}\mathbb{E}\left[\frac{\partial^2 \ln f\left ( \mathbf{Z};\rho \right ) }{\partial \rho^2}\right] \\
&=-N\frac{\partial^2 \phi\left(\rho \right ) }{\partial \rho^2}=-N\sigma_{\eta}^2
\end{align*}
\begin{comment}
Using the property of the log-partition function mentioned in \eqref{p:2}, we have
\begin{equation*}
\frac{\partial^2 \phi\left(\rho \right ) }{\partial \rho^2}=\sigma_{\eta}^2
\end{equation*}
\end{comment}
and the proof readily follows.
\end{IEEEproof}
%\subsubsection{Gaussian Distribution}
%For the particular case when $f\left(\mathbf{U};\xi\right)\sim\mathcal{N}(\xi,\sigma_{\xi}^2)$, it follows from Example 1 that $\sigma_{\eta}^2=\frac{1}{\sigma_{\xi}^2}$, and the CRLB for a Gaussian distributed likelihood can be expressed as
%\begin{equation}
%\mathrm{Var}(\hat\xi)\geq \frac{\sigma_{\xi}^2}{N}
%\end{equation}
%which is a well-known result in estimation theory \cite{kay:est}.
\subsection{Chapman-Robbins Bound}
Chapman-Robbins bound (CHRB), proposed in \cite{chapman:bound}, sets a lower bound on the variance of an estimator of a deterministic parameter. The CHRB does not make any assumptions on the differentiability of the likelihood function and the regularity conditions that often constrain the use of CRB, and is substantially tighter than the CRB in many situations. Hence, CHRB is employed to determine a lower bound on the variance of an unbiased estimator of $\rho$ for constrained likelihood functions.

In general for a parameter $\rho$, the CHRB is given by
\begin{equation}
\mathrm{Var}(\hat\rho)\geq \left [ \underset{h}{\inf}\frac{1}{h^2}\left \{\mathbb{E}\left ( \frac{f(\mathbf{Z};\rho+h)}{f(\mathbf{Z};\rho)} \right )^2-1  \right \} \right ]^{-1}\;,\label{crb}
\end{equation}
which can be evaluated as shown below.
\begin{theorem}
The CHRB for the \label{th:5}parameter $\rho$ given the likelihood function \eqref{U:cons:recall} can be expressed as
\begin{equation}
\mathrm{Var}(\hat\rho)\geq \left [ \underset{h}{\inf}\frac{\left\{\left(M_{\eta}(h)\right)^{-2N}\cdot \zeta^N (h)-1\right\}}{h^2} \right ]^{-1}\label{crb:our}
\end{equation}
where %$\eta(U)\overset{\Delta}{=}\eta(U_j)$ and $U\overset{\Delta}{=}U_j~\forall j=1,\ldots,N$, since $U_j$ are \emph{i.i.d},
$M_{\eta}(h)$ is the MGF of the statistic $\eta(Z_j)$ and
\begin{equation}
\zeta(h)\overset{\Delta}{=}\mathbb{E}\left[\exp\left(2h\eta(Z_j)\right)\mathbb{I}\left(Z_j-\rho-h\right)\right]
\label{zeta:def}
\end{equation}
with the expectation taken with respect to any $Z_j$.
\end{theorem}
\begin{IEEEproof}
The details of the proof are relegated to Appendix \ref{app:CRB}.
\end{IEEEproof}
\subsection{Bayesian Cramer-Rao Lower Bound}
The Bayesian Cramer-Rao bound (BCRB) is a lower bound on the variance of an unbiased estimator when the parameter assumes a prior density. It requires the same regularity conditions to be satisfied as its classical counterpart.

For an estimator $\hat{\rho}_k$ of $\rho_k$, the BCRB states that the variance of the estimator is bounded below by the lower-right sub-matrix of the inverse of the Bayesian information matrix, $ J_{\rm CR}^{-1} (k)$ \cite{van:bayes}, i.e.,
\begin{equation}
\mathrm{Var}\left( \hat{\rho}_k\right) \geq J_{ \rm{CR}}^{-1} (k)  = [ \boldsymbol{J}_{ \rm{CR}}^{-1} (k) ]_{kk}  \label{bcrb}
\end{equation}
where the Bayesian information matrix is given by
\begin{align*}
[ \boldsymbol{J}_{\rm {CR}} (k) ]_{ij} & \overset{\Delta}{=} \mathbb{E} \left[ \frac{\partial \log f(\mathbf{Z}_k, \bm{\rho}_k)}{\partial \rho_i} \frac{\partial \log f(\mathbf{Z}_k, \bm{\rho}_k)}{\partial \rho_j} \right]\nonumber \\
& = -\mathbb{E} \left[ \frac{\partial^2 \log f(\mathbf{Z}_k, \bm{\rho}_k)}{\partial \rho_i \partial \rho_j} \right]
\end{align*}
where the expectation is taken with respect to the joint pdf and
\begin{align}
\mathbf{Z}_k & \overset{\Delta}{=}  [ Z_1, \ldots, Z_k ]^{T} \nonumber \\
\bm{\rho}_k & \overset{\Delta}{=}  [ \rho_0,\rho_1, \ldots, \rho_k ]^{T}\nonumber \\
f(Z_k|\rho_k) & \propto \exp \left( \eta(Z_k) \rho_k - \phi_k(\rho_k) \right)\;.\label{pdf:bayes}
\end{align}
It is assumed that the parameter $\rho_k$ evolves through a Gauss-Markov model given by
\begin{equation}
f(\rho_k | \rho_{k-1}) = \frac{1}{\sqrt{2 \pi\sigma^2 }} \exp \left( - \frac{(\rho_k - \rho_{k-1})^2}{2 \sigma^2} \right)\;. \label{pdf:rho:k}
\end{equation}
A recursive formula to evaluate the Bayesian sub-matrix, derived in \cite{tichav:recursive}, is given by
\begin{equation}
\begin{split}
J_{\rm {CR}}(k+1) =& - E_{\rm {CR}}^{(2)}(k) \left( J_{\rm {CR}} (k) + E_{\rm {CR}}^{(1)}(k) \right)^{-1} E_{\rm {CR}}^{(2)}(k) \\
& +  E_{\rm {CR}}^{(3A)}(k) + E_{\rm {CR}}^{(3B)}(k)\label{bound:recursive}
\end{split}
\end{equation}
where
\begin{align*}
E_{\rm {CR}}^{(1)}(k) & \overset{\Delta}{=}  \mathbb{E} \left[ - \frac{\partial^2 }{\partial \rho_{k}^2} \log f(\rho_{k+1} | \rho_{k} ) \right] \\
E_{\rm {CR}}^{(2)}(k)& \overset{\Delta}{=}  \mathbb{E} \left[ - \frac{\partial^2 }{\partial \rho_k \partial \rho_{k+1}} \log f(\rho_{k+1} | \rho_{k} ) \right]\\
E_{\rm {CR}}^{(3A)}(k) & \overset{\Delta}{=}  \mathbb{E} \left[ - \frac{\partial^2 }{\partial \rho_{k+1}^2} \log f(\rho_{k+1} | \rho_{k} ) \right]\\
E_{\rm {CR}}^{(3B)}(k) & \overset{\Delta}{=}  \mathbb{E} \left[ - \frac{\partial^2}{\partial \rho_{k+1}^2} \log f(Z_{k+1} | \rho_{k+1} ) \right]
\end{align*}
and the expectation is again with respect to the joint pdf.\\
\begin{theorem}
For the Bayesian framework in \eqref{pdf:bayes} and \eqref{pdf:rho:k}, the recursive Bayesian information matrix in \eqref{bound:recursive} is given by
\begin{equation}
J_{\rm {CR}}(k+1) = \left( \sigma^2 + J^{-1}_{\rm {CR}}(k) \right)^{-1} + \sigma_{\eta_{k}}^2\label{bayes:crlb:our}
\end{equation}
with $J_{\rm {CR}}(0) = 0$.
\end{theorem}
\begin{IEEEproof}
For the density functions, $f(\rho_k|\rho_{k-1})$ and $f(Z_k|\rho_k)$ in \eqref{pdf:bayes}, it can be verified that
\begin{equation*}
E_{\rm {CR}}^{(1)}(k)= \frac{1}{\sigma^2}\;, \quad E_{\rm {CR}}^{(2)}(k) = - \frac{1}{\sigma^2}\;, \quad E_{\rm {CR}}^{(3A)}(k) = \frac{1}{\sigma^2}
\end{equation*}
and
\begin{align*}
E_{\rm {CR}}^{(3B)}(k) &= \int \int \frac{\partial^2 \phi_k(\rho_{k+1})}{\partial \rho_{k+1}^2} f(\rho_{k+1}, Z_{k+1}) d \rho_{k+1} d Z_{k+1} \\
&= \frac{\partial^2 \phi_k(\rho_{k+1})}{\partial \rho_{k+1}^2} = \sigma_{\eta_{k}}^2\;.
\end{align*}
The proof follows by plugging these quantities in \eqref{bound:recursive}.
\end{IEEEproof}
\subsection{Bayesian Chapman-Robbins Bound}
A Bayesian version of the Chapman-Robbins bound (BCHRB) can be used to provide a lower bound on the variance of an estimator of $\rho_k$ when there are no regularity assumptions on the likelihood. In fact, unlike the BCRB, the BCHRB can be evaluated for constrained likelihood functions where the domain of the likelihood is dependent on the unknown parameter.

BCHRB states that the variance of an estimator $\bm{\hat{\rho}}_k$ of $\bm{\rho}_k$ is lower bounded as
 %The statement of the bound is ($\mathbf{h}_k \overset{\Delta}{=} [h_o, h_1, \ldots, h_k]^{T}$)
\begin{equation*}
\mathrm{Var} (\bm{\hat{\rho}}_k) - \left[ T_k(\mathbf{h}_k) - 1 \right]^{-1} \mathbf{h}_k \mathbf{h}_k^T \succeq  \mathbf{0}
\end{equation*}
with $\succeq $ in the positive semi-definite sense, where
\begin{equation*}
T_k(\mathbf{h}_k) \overset{\Delta}{=} \mathbb{E}\left[ \left( \frac{f(\mathbf{Z}_k, \bm{\rho}_k + \mathbf{h}_k)}{f(\mathbf{Z}_k, \bm{\rho}_k)} \right)^2 \right] \;,
\end{equation*}
and $\mathbf{h}_k  \overset{\Delta}{=}  [0, h_1, \ldots, h_k ]^{T}$.\\
\begin{theorem}
The BCHRB \label{th:7}for the parameter $\rho_k$ can be expressed as
\begin{equation*}
\mathrm{Var} (\hat{\rho}_k)\geq \frac{1}{J_{{\textrm {CH}}, k}}
\end{equation*}
where
\begin{equation*}
J_{{\textrm {CH}}, k} = \inf_{\mathbf{h}_k} \frac{T_k(\mathbf{h}_k)-1}{h_k^2}
\end{equation*}
\end{theorem}
and
\begin{equation}
\begin{split}
T_k(\mathbf{h}_k) =& \left( \prod_{j = 1}^{k} M^{-2}_{\eta} (h_j) M_{\eta} (2h_j) \right) \times \\
&  \exp \left[ \frac{1}{\sigma^2} \sum_{j=1}^k \left( h_j - h_{j-1} \right)^2 \right]\;.
\end{split}
\label{CHR_general_T_k}
\end{equation}
\begin{IEEEproof}
See Appendix \ref{app:BCHBgeneral} for details.
\end{IEEEproof}\vspace{0.15in}
\begin{remark}
Since the performance bounds are derived for arbitrary exponential family distributions, they can also prove useful in a broad sense in classical as well as Bayesian parameter estimation.
\end{remark}

\subsection{Relation to Clock Offset Estimation}
The performance bounds derived in the preceding subsections can be used to lower bound the mean square error (MSE) of the clock offset $\theta$. Notice the similarity between the unconstrained and constrained likelihood
functions for $\xi$ and $\psi$ in \eqref{U:uncons}-\eqref{V:cons} and the general exponential family likelihood function considered in \eqref{U:uncons:recall} and \eqref{U:cons:recall}. Therefore, the bounds derived above for
 $\rho$ are also applicable to the parameter $\xi$ (and also $\psi$). The MSE of the clock offset $\theta$ can, in turn, be lower bounded using the bounds on $\xi$ and $\psi$.

Using \eqref{subs:theta}, the following result is immediate.
\begin{proposition}
The MSE\label{prop:1} of any estimator of $\theta$ can be expressed as
\begin{equation*}
\mathrm{MSE}  \left( \hat{\theta} \right) = \frac{1}{4} \left( \textrm{Var}  \left( \hat{\xi} \right) +
\mathrm{Var}  \left( \hat{\psi} \right) \right) + \frac{1}{4} \left( b_{\xi} - b_{\psi} \right)^2
\end{equation*}
where $b_{\xi}$ and $b_{\psi}$ are the biases of the estimators $\hat\xi$ and $\hat\psi$, respectively.
\end{proposition}
The explicit lower bounds on the MSE of any estimator $\hat\theta$ for classical as well as Bayesian framework in case of Gaussian and exponentially distributed likelihood functions can be evaluated as shown below.

\subsubsection{Gaussian Distribution - CRB}
If the likelihood function for $\xi$ is Gaussian distributed \eqref{example:gauss:pdf}, then using \eqref{example:gauss:param} and \eqref{crlb:our}, it is straightforward to see that the CRB for any unbiased estimator $\hat{\xi}$ is given by
\begin{equation*}
\mathrm{Var} \left( \hat{\xi} \right) \geq \frac{\sigma_{\xi}^2}{N} \;,
\end{equation*}
and a similar expression is applicable to $\hat{\psi}$ as well. Using Proposition \ref{prop:1}, it can be concluded that
\begin{equation}
\mathrm{MSE}  \left( \hat{\theta} \right) \geq \frac{\sigma_{\xi}^2 + \sigma_{\psi}^2}{4N} \;.
\label{CRB_MLE_gaussian}
\end{equation}
As a remark, it is evident in this case that $\hat{\theta}_{\textrm{ML}}$ \eqref{example:gauss:theta:ML} is efficient in the sense that its MSE achieves \eqref{CRB_MLE_gaussian} with equality (cf. Appendix \ref{app:MSE_bounds_classical_gauss}).

\subsubsection{Exponential Distribution - CHRB}
If the likelihood for $\xi$ is exponentially distributed \eqref{example:exp:pdf}, using \eqref{mgf} and \eqref{example:exp:param}, it can be easily verified that
\begin{equation*}
M_{\eta_{\xi}(U)}(h) = 1
\end{equation*}
and \eqref{zeta:def} becomes
\begin{equation*}
\zeta (h) = \exp \left( \lambda_{\xi} h \right) \;,
\end{equation*}
so that the statement of the CHRB \eqref{crb:our} can be rewritten as
\begin{equation*}
\mathrm{Var}  \left( \hat{\xi} \right) \geq \left[ \inf_{h} \frac{\exp \left( \lambda_{\xi} h N \right) - 1 }{h^2} \right]^{-1} = \frac{0.6476}{\lambda_{\xi}^2 N^2}
\end{equation*}
and similarly for $\hat{\psi}$. Using Proposition \ref{prop:1}, it follows that
\begin{equation}
\begin{split}
\mathrm{MSE}  \left( \hat{\theta} \right) &= \frac{1}{4} \left( \textrm{Var}  \left( \hat{\xi} \right) + \mathrm{Var}  \left( \hat{\psi} \right) \right) + \frac{1}{4} \left( b_{\xi} - b_{\psi} \right)^2 \\
&\geq \frac{0.162}{N^2} \left( \frac{1}{\lambda_{\xi}^2} + \frac{1}{\lambda_{\psi}^2} \right) + \frac{1}{4} \left( b_{\xi} - b_{\psi} \right)^2 \;.
\end{split}
\label{CHB_MLE_exp_distr}
\end{equation}

\subsubsection{Gaussian Distribution - BCRB}
In the Bayesian regime, if the likelihood function for $\xi$ is Gaussian distributed \eqref{pdf:gauss:k}, by using \eqref{gauss:param:k} and \eqref{bayes:crlb:our}, it can be seen that
\begin{equation*}
J_{\textrm{CR}, \xi} \left( k + 1 \right) = \left( \sigma^2 + J^{-1}_{\textrm{CR}, \xi} \left( k \right) \right)^{-1} + \frac{1}{\sigma_{\xi,k}^2} \;,
\end{equation*}
with $J_{\textrm{CR}, \xi} \left( 0 \right) = 0$. A similar line of reasoning can be followed to derive an analogous recursion for $J_{\textrm{CR}, \psi} \left( k \right)$. The MSE of $\theta$ can be now be lower bounded as
\begin{equation}
\mathrm{Var}(\hat\theta_k)\geq \frac{1}{4}\left(\frac{1}{J_{\textrm{CR}, \xi} \left( k \right)}+\frac{1}{J_{\textrm{CR}, \psi} \left( k \right)}\right)\;.
\end{equation}

\subsubsection{Exponential Distribution - BCHRB}
If the likelihood for $\xi_k$ is exponentially distributed \eqref{U:exp}, \eqref{CHR_general_T_k} turns out to be
\begin{equation*}
T_k(\mathbf{h}_k) = \exp \left( \lambda_{\xi} \sum_{j = 1}^{N} h_j \right) \exp \left[ \frac{1}{\sigma^2} \sum_{j=1}^k \left( h_j - h_{j-1} \right)^2 \right] \;.
\end{equation*}
In fact, we just have to notice that $\phi_{\xi} \left( \xi_k \right)$ is a constant function over $\xi_k$ and $\eta_{\xi}(U_j) = \lambda_{\xi}$, so that \eqref{exp_2_h_eta_general} becomes
\begin{equation*}
\mathbb{E} \left[ \exp \left( 2 h_j \eta_{\xi}(U_j) \right) \right] = \exp \left( \lambda_{\xi} h_j \right)
\end{equation*}
therefore $S(\mathbf{h}_k) = \exp \left( \lambda_{\xi} \sum_{j = 1}^{N} h_j \right)$.

\section{Simulation Results}
This \label{sec:simul}section aims to corroborate the theoretical results in preceding sections by conducting simulation studies in various scenarios. The performance of both the classical as well as Bayesian estimators is to be investigated. The measure of fidelity used to rate this performance is the MSE of the estimators for $\theta$ and $\theta_N$. The parameter choice is $\sigma_{\xi} = \sigma_{\psi} = 0.1$ for both Gaussian and log-normally distributed likelihoods, while $\lambda_{\xi} = \lambda_{\psi} = 10$ for exponentially distributed likelihood functions.

\subsection{Log-normal Distribution}
No solution is reported thus far in literature in case the likelihood is log-normally distributed. The proposed estimators, $\hat\theta_{\textrm{ML}}$ \eqref{example:lognormal:theta:ML} and $\hat\theta_N$ \eqref{FGE_theta_logn}, can be used to determine an estimate for the clock offset in classical and Bayesian framework, respectively, as shown below.

\subsubsection{Classical Estimation Framework}
\begin{figure}
\centering
\includegraphics[width=1\hsize]{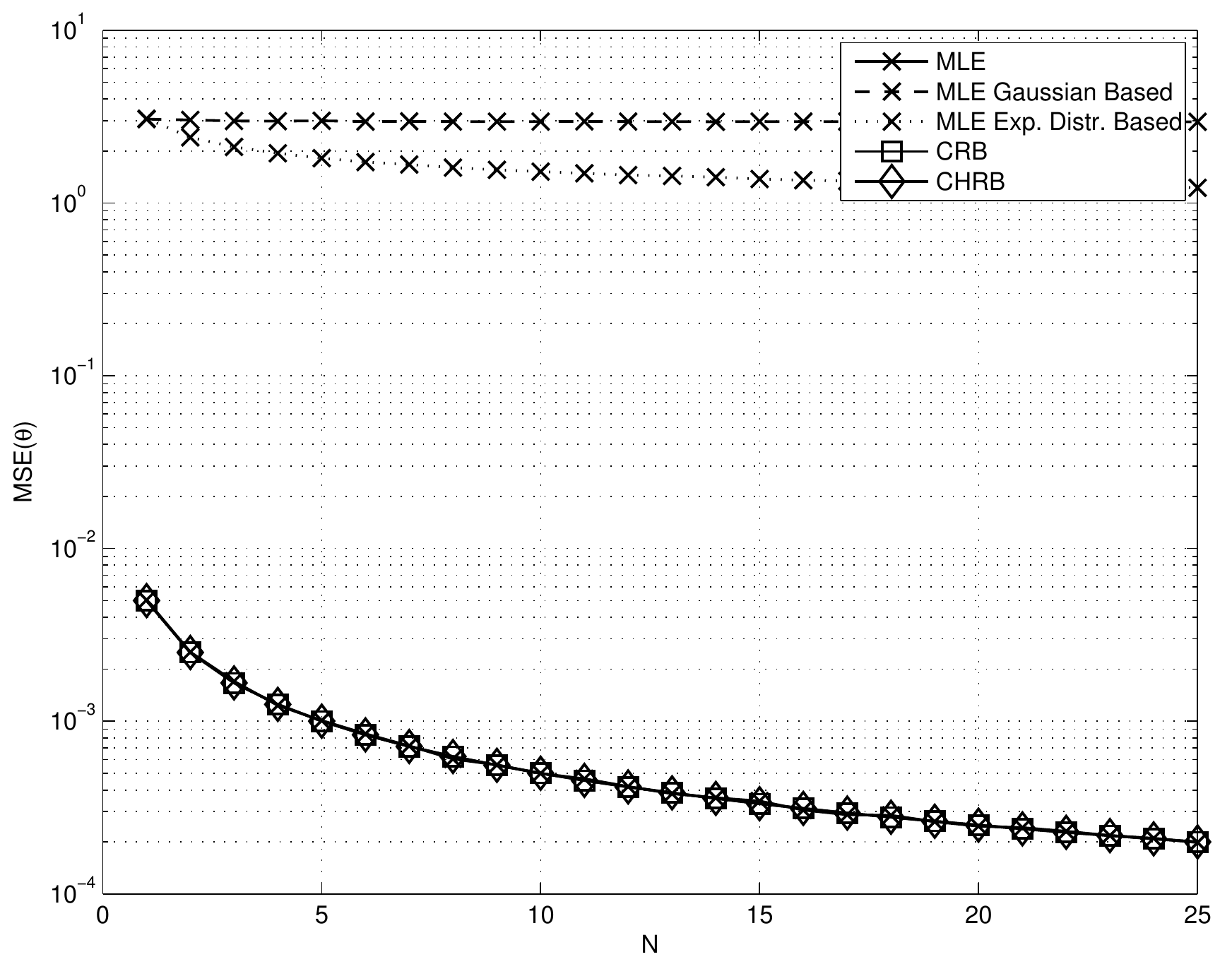}
\caption{MSE and bounds for estimating $\theta$ by using the MLE with log-normal likelihood.}
\label{ML_theta_vs_k_lognormal}
\end{figure}

The existing approaches in literature only consider the Gaussian and the exponential cases, therefore \eqref{example:lognormal:theta:ML}  is a new result in the state-of-the-art about clock offset estimation.
\begin{comment}Before now, if in reality the likelihood of the readings was log-normally distributed, suboptimal estimators have been used, which were assuming the likelihood to be either Gaussian or exponentially distributed.\end{comment}
Fig. \ref{ML_theta_vs_k_lognormal} shows a comparison between the proposed ML estimator (MLE) \eqref{example:lognormal:theta:ML} in case of a log-normally distributed likelihood \eqref{example:lognormal:pdf} with MLEs which (wrongly) assume that the likelihood is Gaussian and exponentially distributed, respectively. The plot shows that the latter approaches are not robust with respect to the likelihood distribution, and their performance is extremely poor if their assumptions do not hold. In addition, Fig. \ref{ML_theta_vs_k_lognormal} also shows that the proposed MLE \eqref{example:lognormal:theta:ML} is efficient since it attains the CRB (as well as the CHRB).

\subsubsection{Bayesian Estimation Framework}
\begin{figure}
\centering
\includegraphics[width=1\hsize]{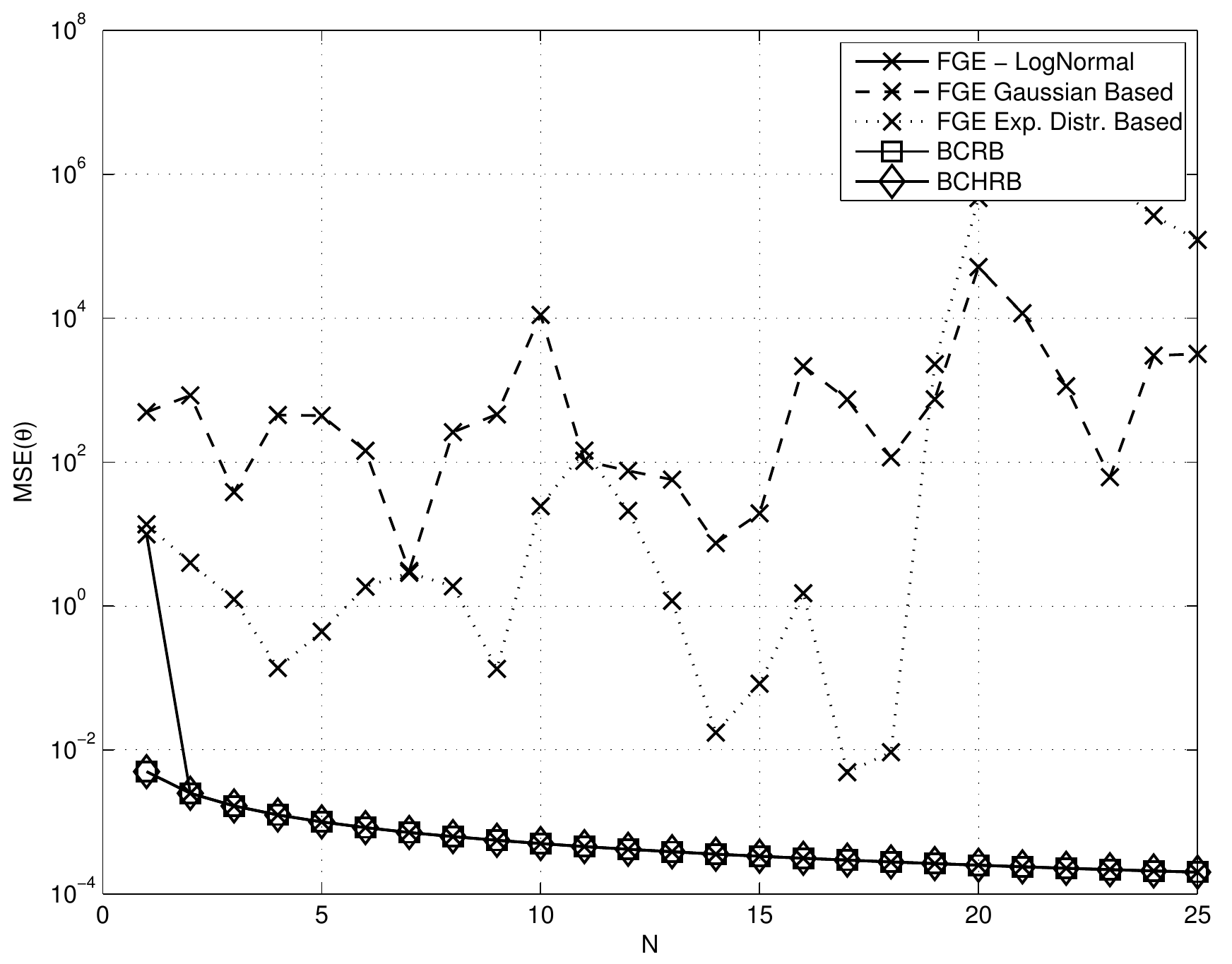}
\caption{MSE and bounds for estimating $\theta_N$ by using FGE with log-normal likelihood.}
\label{Bayes_theta_eps_lognormal}
\end{figure}

%\begin{figure}
%\centering
%\includegraphics[width=1\hsize]{images/Bayes_theta_vs_k_lognormal.eps}
%\caption{MSE and bounds for estimation of $\theta_k$ in the Bayesian framework with log-normal likelihood.}
%\label{Bayes_theta_eps_lognormal}
%\end{figure}
Fig. \ref{Bayes_theta_eps_lognormal} plots the MSE performance of the FGE \eqref{FGE_theta_logn} as well as the BCRB and BCHRB when the likelihoods are log-normally distributed \eqref{exampleb:lognormal:pdf}, and $\sigma = 10^{-4}$. Firstly, it can be seen that the MSE of the proposed FGE coincides with the estimation bounds. Secondly, as in the classical estimation case, if we were to (wrongly) assume a Gaussian or exponential distribution for the likelihoods \eqref{pdf:uncons:bayes}, the resulting FGEs would perform poorly, a fact is evident in Fig. \ref{Bayes_theta_eps_lognormal} by observing the unboundedness and unpredictability of the dashed curve (Gaussian assumption for the likelihoods) and the dotted curve (likelihoods assumed exponentially distributed). This clearly establishes that the FGE \eqref{FGE_theta_logn}, obtained assuming log-normally distributed likelihoods, allows a strong performance improvement with respect to existing estimators if the likelihood functions \eqref{pdf:uncons:bayes} are actually log-normally distributed.

\subsection{Estimator Performance vs Estimation Bounds}
It will also be useful to asses the performance of the MLEs in Gaussian and exponential cases derived in Section \ref{sec:ml} against the various benchmark estimation bounds derived in Section \ref{sec:bounds}. Similarly, the FGEs for Gaussian and exponential distributions, proposed in Section \ref{sec:factor}, can also be compared with the Bayesian bounds to study their MSE performance.

\subsubsection{Classical Estimation Framework}
\begin{figure}
\centering
\includegraphics[width=1\hsize]{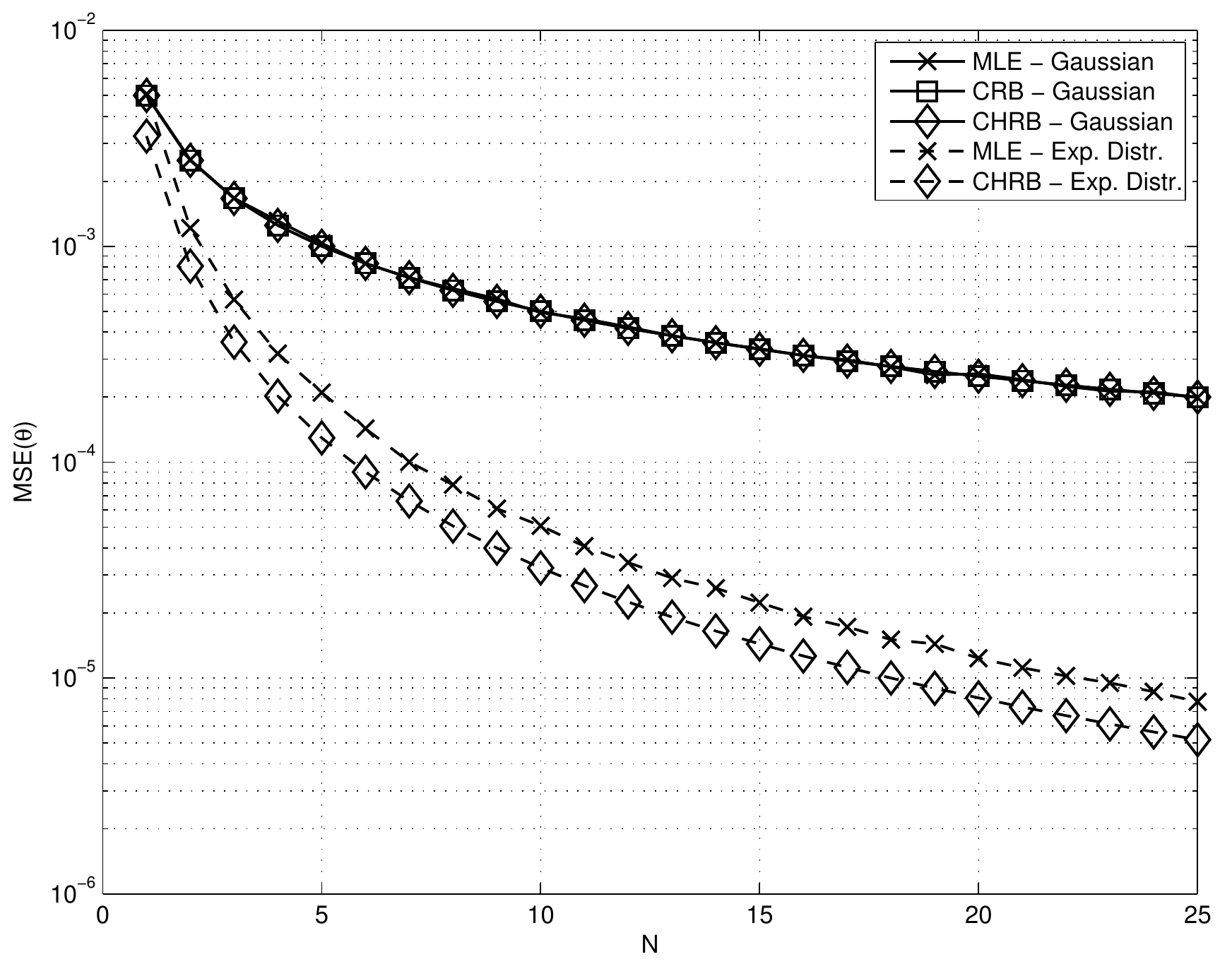}
\caption{MSE and bounds for estimating $\theta$ by using the MLE with Gaussian and exponentially distributed likelihood.}
\label{ML_theta_eps}
\end{figure}

Fig. \ref{ML_theta_eps} shows the performance comparison between the MSE of the MLEs \eqref{example:gauss:theta:ML} and \eqref{example:exp:theta:ML} for Gaussian and exponentially distributed likelihood functions against the CRB and the CHRB, respectively. Firstly, it is evident that in the case of Gaussian distribution, the CRB and the CHRB coincide. Moreover, the MSE of $\hat{\theta}_{\textrm{ML}}$ also coincides with the aforementioned bounds. On the other hand, for an exponentially distributed likelihood function, due to its lack of regularity, the CRB cannot be derived, thus only the CHRB is shown. It can be observed that the MSE of $\hat{\theta}_{\textrm{ML}}$ is fairly close to CHRB, even though it does not coincide with it. From Fig. \ref{ML_theta_eps} the MSE of the MLEs for the Gaussian and exponential distribution case can be also compared. In order to ensure a fair comparison, parameters are chosen in a way to have the same variance of the observations for both distributions. From the MSE curves, one can infer that the MSE in case of an exponentially distributed likelihood is lower than the one for a Gaussian distribution as the number of observations $N$ increases. This behavior is expected since it can be verified by the MSE expressions \eqref{MSE_MLE_gaussian} and \eqref{MSE_MLE_exp_distr}, reported in Appendix \ref{app:MSE_bounds_classical}, that in case of a Gaussian distribution, the MSE decreases proportionally to $1/N$, while in the exponential distribution case it decreases proportionally to $1/N^2$.
\begin{comment}In Appendix \ref{app:MSE_bounds_classical} the expressions for the main bounds (CRB for the Gaussian case \eqref{CRB_MLE_gaussian} and CHRB for the exponentially distributed likelihoods \eqref{CHB_MLE_exp_distr} are also reported, for the sake of completeness.
\end{comment}

\subsubsection{Bayesian Estimation Framework}
\begin{figure}
\centering
\includegraphics[width=1\hsize]{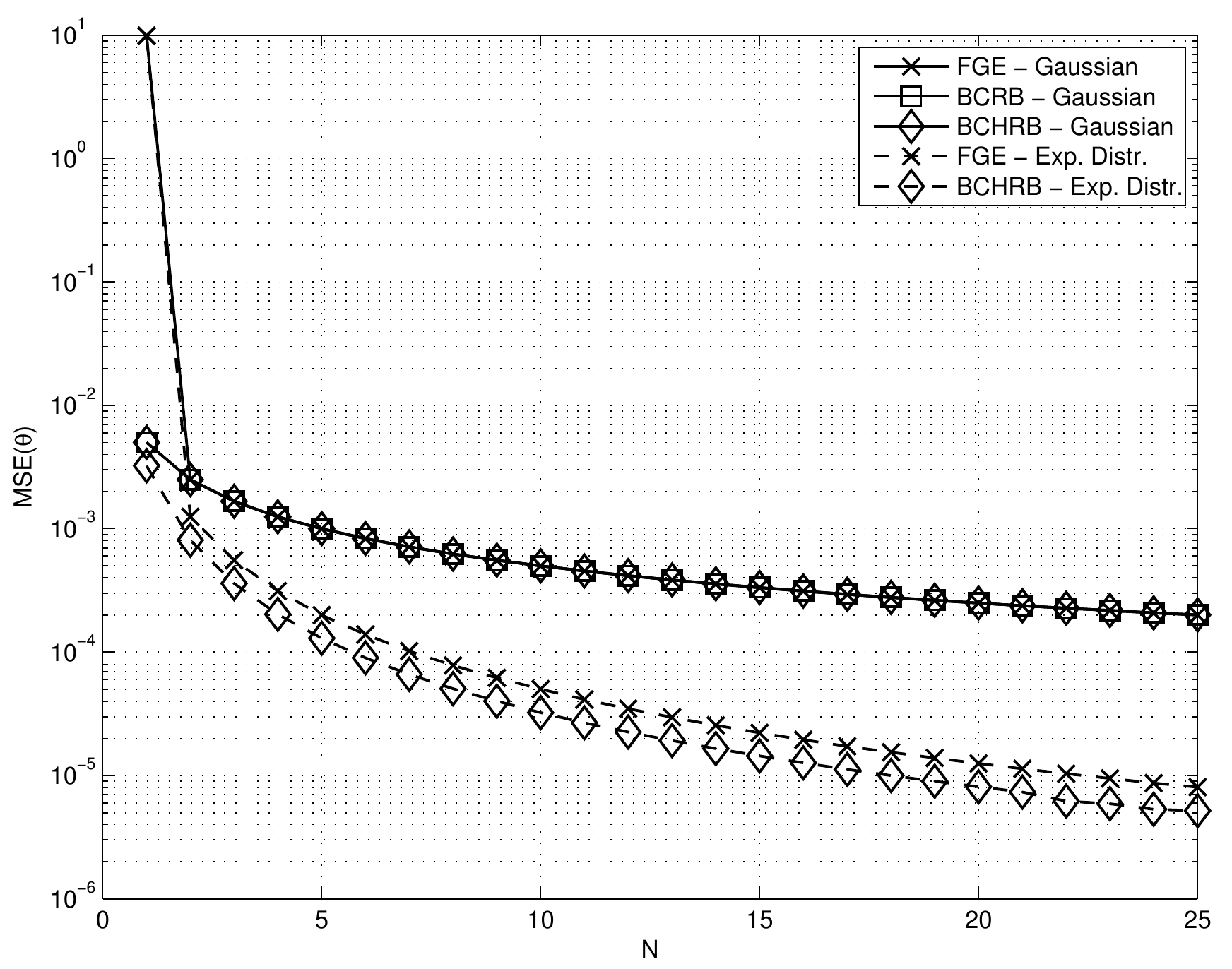}
\caption{MSE and bounds for estimating $\theta_N$ by using FGE with Gaussian and exponentially distributed likelihood.}
\label{Bayes_theta_eps}
\end{figure}

In Fig. \ref{Bayes_theta_eps}, the MSE performance of the FGEs $\hat\theta_N$ \eqref{FGE_theta_gauss} and \eqref{example:exp:theta:N} is compared with BCRB and BCHRB for $\sigma = 10^{-4}$. As in the classical estimation scenario, it is evident that for Gaussian distributed likelihoods, the MSE using \eqref{FGE_theta_gauss} for $\hat{\theta}_N$ coincides with the reported bounds. The MSE of the FGE derived assuming exponentially distributed likelihoods \eqref{example:exp:theta:N} is plotted against the BCHRB as well in Fig. \ref{Bayes_theta_eps}. It is clear that the MSE is quite close to BCHRB, although not coinciding with it, as exactly was the case in the classical estimation framework.

\subsection{Comparing Classical and Bayesian frameworks}
\begin{figure}
\centering
\includegraphics[width=1\hsize]{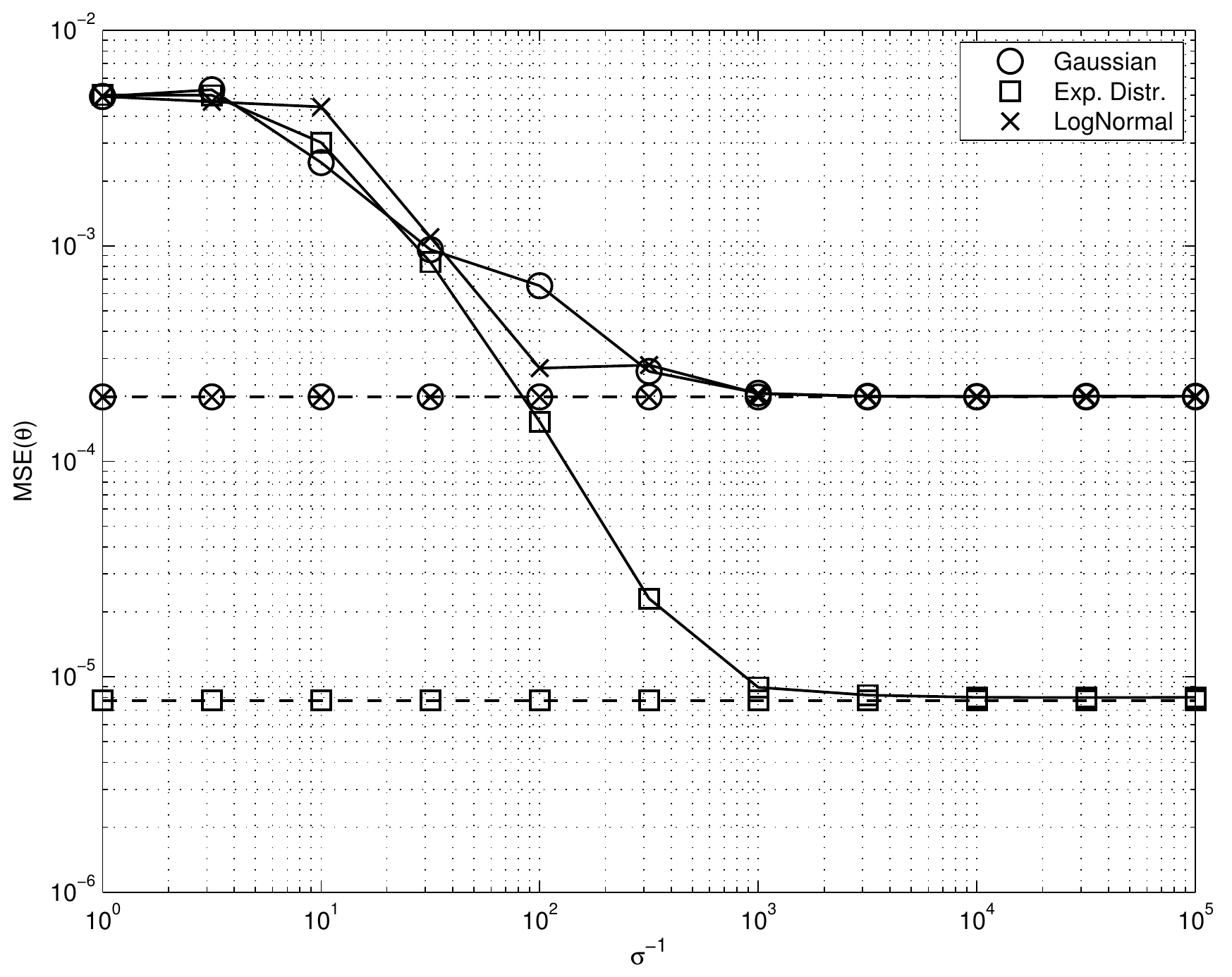}
\caption{MSE in estimating of $\theta_N$ vs $\sigma$. Solid lines: FGE; dashed lines: MLE (classical framework).}
\label{Bayes_theta_vs_sigma_GM}
\end{figure}
The estimators proposed in the classical and the Bayesian framework can also be compared with each other based on their MSE performance as the system noise decreases. The aim here is to show that the latter approaches the former as $\sigma \to 0$.

Fig. \ref{Bayes_theta_vs_sigma_GM} depicts the MSE for the cases of Gaussian, exponential and log-normal distribution for the likelihoods with $N = 25$. In the plot, the horizontal lines represent the MSEs in the classical framework, obtained with the MLEs, as shown in \eqref{MSE_MLE_gaussian} and \eqref{MSE_MLE_exp_distr} in Appendix \ref{app:MSE_bounds_classical}. It can be observed that, for all the three considered distributions, the MSE obtained by using the FGE for estimating $\theta$ approaches the MSE of the MLEs as $\sigma < 10^{-3}$. %Moreover, it is clear from the curves that for the Gaussian and log-normal likelihood, the MLE performance is approached before (higher values of $\sigma^2$) with respect to the case of an exponentially distributed likelihood with the same variance.

\section{Conclusions and Future Work}
The clock synchronization problem in sensor networks has received keen attention recently owing to its central role in critical network operations as duty cycling, data fusion, and node localization. Based on a two-way timing message exchange scenario, this work proposes a unified framework for the clock offset estimation problem when the likelihood function of the observation time stamps is Gaussian, exponential and log-normally distributed. A convex optimization based ML estimation approach is presented for clock offsets. The results known thus far for Gaussian and exponentially distributed network delays are subsumed in the general approach while the ML estimator is derived when the likelihood function is log-normally distributed. In order to study the case of a possibly time-varying clock offset, a Bayesian approach is also studied using factor graphs. The novel message passing strategy results in a closed form solution of the time-varying clock offset estimation problem. In order to compare various estimators, several lower bounds on the variance of an estimator have been derived in the classical as well as the Bayesian regime for likelihood functions which are arbitrary members of the exponential family, a wide class containing several distributions of interest. The theoretical findings are corroborated by simulation studies conducted in various scenarios.

In future, it will be useful to incorporate the effect of clock skew in the clock offset estimation model. This can result in further reduction of the re-synchronization periods. In addition, the results about pairwise synchronization can be used to build a framework for network-wide synchronization across a sensor network. \label{sec:concl}

\appendices
\begin{comment}
\section{Proof of Lemma 2}
We \label{app:lemma}have
\begin{align}
&\mathbb{E}\left[\left(\hat\theta_{AML}-\theta\right)^2\right]=\mathbb{E}\left[\left(\frac{\hat\xi_{AML}-
\hat\psi_{AML}}{2}-\frac{\xi-\psi}{2}\right)^2\right]\nonumber \\
&=\frac{\mathbb{E}\left[(\hat\xi_{AML}-\hat\psi_{AML}-\xi+\psi)^2\right]}{4}\nonumber \\
&=\frac{1}{4}\mathbb{E}\left[(\hat\xi_{AML}-\xi)^2\right]+\frac{1}{4}\mathbb{E}\left[(\hat\psi_{AML}-\psi)^2\right]\nonumber\\
&-\frac{1}{2}\mathbb{E}\left[(\hat\xi_{AML}-\xi)(\hat\psi_{AML}-\psi)\right]\nonumber \\
&=\frac{1}{4}\mathrm{MSE}(\hat\xi_{AML})+\frac{1}{4}\mathrm{MSE}(\hat\psi_{AML})\nonumber \\
&-\frac{1}{2}\mathbb{E}\left[(\hat\xi_{AML}-\xi)\right]\mathbb{E}\left[(\hat\psi_{AML}-\psi)\right]\label{mse:lemma2}
\end{align}
where \eqref{mse:lemma2} follows from the independence of $\hat\xi_{AML}$ and $\hat\psi_{AML}$. Using \eqref{xi:psi:ML} from Theorem \ref{th:1}, the first expectation in the product can be evaluated as
\begin{align}
\mathbb{E}\left[(\hat\xi_{AML}-\xi)\right]&=\mathbb{E}\left[\frac{\sum_{j=1}^{N}\eta_{\xi}(U_j)}{N\sigma_\eta^2}-\xi
\right]\\
&=\frac{\sum_{j=1}^{N}\mathbb{E}\left[\eta_{\xi}(U_j)\right]}{N\sigma_\eta^2}-\xi\nonumber \\
&=\sum_{j=1}^{N}\frac{\frac{\partial \phi_{\xi}(\xi)}{\partial \xi}}{N\sigma_\eta^2}-\xi\nonumber \\
&=\sum_{j=1}^{N}\frac{2a\xi}{N\sigma_\eta^2}-\xi\nonumber \\
&=0
\end{align}
where we have used \eqref{p:1}, \eqref{p:2} and \eqref{lpf:approx}. This implies that the estimators $\hat\xi_{AML}$ and $\hat\psi_{AML}$ are unbiased and the proof follows.
\end{comment}

\section{Proof of Theorem \ref{th:5}}
The ratio of\label{app:CRB} the likelihood functions can be expressed as
\begin{align*}
&\frac{f(\mathbf{Z};\rho+h)}{f(\mathbf{Z};\rho)}\\
\hspace{-5mm}=& \frac{e^{\left((\rho+h)\sum_{j=1}^{N}\eta(Z_j)-N\phi(\rho+h)\right)}\prod_{j=1}^{N}\mathbb{I}(Z_j-\rho-h)}{ e^{\left(\rho\sum_{j=1}^{N}\eta(Z_j)-N\phi(\rho)\right)\prod_{j=1}^{N}\mathbb{I}(Z_j-\rho)}}\\
=&e^{\left(h\sum_{j=1}^{N}\eta(Z_j)-N\phi(\rho+h)+N\phi(\rho)\right)}\prod_{j=1}^{N}\mathbb{I}(Z_j-\rho-h) \\
=&e^{\left(h\sum_{j=1}^{N}\eta(Z_j)\right)}e^{\left(-N(\phi(\rho+h)+\phi(\rho))\right)}
\prod_{j=1}^{N}\mathbb{I}(Z_j-\rho-h)\;.
\end{align*}
%\begin{align}
%\frac{f(\mathbf{U};\xi+h)}{f(\mathbf{U};\xi)}=& \frac{\exp\left((\xi+h)\eta'-N\phi(\xi+h)\right)}{ \exp\left(\xi\eta'-N\phi_{\xi}(\xi)\right)}\nonumber\\
%=\exp&\left(h\eta'-N\phi(\xi+h)+N\phi_{\xi}(\xi)\right)\nonumber \\
%=\exp&\left(h\eta'\right)\exp\left(-N(\phi(\xi+h)+\phi_{\xi}(\xi))\right)
%\end{align}
The expectation of the ratio of the likelihood functions can now be calculated as
\begin{align*}
&\mathbb{E}\left ( \frac{f(\mathbf{Z};\rho+h)}{f(\mathbf{Z};\rho)} \right )^2\\
&=\mathbb{E}\left[e^{\left(2h\sum_{j=1}^{N}\eta(Z_j)\right)}e^{\left(-2N(\phi(\rho+h)+\phi(\rho))\right)}
\prod_{j=1}^{N}\mathbb{I}(Z_j-\rho-h)\right]\\
&=e^{\left(-2N(\phi(\rho+h)+\phi(\rho))\right)}\mathbb{E}\left[e^{\left(2h\sum_{j=1}^{N}\eta(Z_j)\right)}
\prod_{j=1}^{N}\mathbb{I}(Z_j-\rho-h)\right]\\
&=\left(M_{\eta(Z)}(h)\right)^{-2N}\mathbb{E}\left[e^{\left(2h\sum_{j=1}^{N}\eta(Z_j)\right)}
\prod_{j=1}^{N}\mathbb{I}(Z_j-\rho-h)\right]
\end{align*}
where it follows from \eqref{mgf} that
\begin{equation*}
\left(M_{\eta(Z)}(h)\right)^{-2N}=e^{-2N\left(\phi(\rho+h)-\phi(\rho)\right)}\;.
\end{equation*}
Since the samples $Z_j$ are \emph{i.i.d},
\begin{align*}
\mathbb{E}\left[e^{\left(2h\sum_{j=1}^{N}\eta(Z_j)\right)}
\prod_{j=1}^{N}\mathbb{I}(Z_j-\rho-h)\right]\\
=\left(\mathbb{E}\left[e^{\left(2h\eta(Z_j)\right)}
\mathbb{I}(Z_j-\rho-h)\right]\right)^N\;.
\end{align*}
With $\zeta(.)$ defined in the theorem, the proof is complete.
\section{Proof of Theorem \ref{th:7}}
We have\label{app:BCHBgeneral}
\begin{equation*}
\begin{split}
T_k(\mathbf{h}_k) \overset{\Delta}{=} & \mathbb{E}\left[ \left( \frac{f(\mathbf{Z}_k, \bm{\rho}_k + \mathbf{h}_k)}{f(\mathbf{Z}_k, \bm{\rho}_k)} \right)^2 \right]\\
= \int_{- \infty}^{+ \infty} \int_{- \infty}^{+ \infty} & \left( \frac{f(\mathbf{Z}_k, \bm{\rho}_k + \mathbf{h}_k)}{f(\mathbf{Z}_k, \bm{\rho}_k)} \right)^2 f(\mathbf{Z}_k, \bm{\rho}_k)  d \mathbf{Z}_k d\bm{\rho}_k  \\
= S(\mathbf{h}_k) \int_{- \infty}^{+ \infty} & \frac{f(\bm{\rho}_k + \mathbf{h}_k)^2}{f(\bm{\rho}_k)} d \bm{\rho}_k
\end{split}
\end{equation*}
where
\begin{equation}
\begin{split}
 S(\mathbf{h}_k) \overset{\Delta}{=} & \int_{- \infty}^{+ \infty}\left( \frac{f(\mathbf{Z}_k | \bm{\rho}_k + \mathbf{h}_k)}{f(\mathbf{Z}_k | \bm{\rho}_k)} \right)^2 f(\mathbf{Z}_k | \bm{\rho}_k) d \mathbf{Z}_k \;.
\end{split}
\label{S_h_def}
\end{equation}
Continuing with the calculations
\begin{equation*}
\begin{split}
T_h(\mathbf{h}_k) =& S(\mathbf{h}_k) \int_{- \infty}^{+ \infty}  \frac{f(\bm{\rho}_k + \mathbf{h}_k)^2}{f(\bm{\rho}_k)} d \bm{\rho}_k \\
=& S(\mathbf{h}_k) \int_{- \infty}^{+ \infty} \frac{f^2(\rho_0 + h_0)}{f(\rho_0)} \times \\
&   \prod_{j = 1}^k \frac{f^2(\rho_j + h_j | \rho_{j-1} + h_{j-1})}{f(\rho_j | \rho_{j-1})} d \bm{\rho}_k \\
=& S(\mathbf{h}_k) \int_{- \infty}^{+ \infty}  \prod_{j = 1}^k \frac{f^2(\rho_j + h_j | \rho_{j-1} + h_{j-1})}{f(\rho_j | \rho_{j-1})} d \bm{\rho}_k \;.
\end{split}
\end{equation*}
Since $\frac{f^2(\rho_j + h_j | \rho_{j-1} + h_{j-1})}{f(\rho_j | \rho_{j-1})}$ can be verified to be equal to ($j = 1, \ldots, k$)
\begin{equation*}
\begin{split}
& \frac{1}{\sigma \sqrt{2 \pi}} \exp \left[ -\frac{\rho_{j-1}^2}{2 \sigma^2} + \frac{\rho_j + 2(h_j - h_{j-1})}{\sigma^2} \rho_{j-1} \right] \times \\
&  \exp \left[ - \frac{(h_j - h_{j-1})^2}{\sigma^2} \right] \exp \left[ -\frac{\rho_j^2}{2 \sigma^2} - \frac{2\rho_j(h_j - h_{j-1})}{\sigma^2} \right]
\end{split}
\end{equation*}
it turns out that
\begin{equation*}
\begin{split}
\int_{- \infty}^{+\infty} \frac{f^2(\rho_j + h_j | \rho_{j-1} + h_{j-1})}{f(\rho_j | \rho_{j-1})} d \rho_{j-1} = \exp \left[ \frac{(h_j - h_{j-1})^2}{\sigma^2} \right]  \;
\end{split}
\end{equation*}
that brings to
\begin{equation*}
T_k(\mathbf{h}_k) = S(\mathbf{h}_k) \exp \left[ \frac{1}{\sigma^2} \sum_{j=1}^k \left( h_j - h_{j-1} \right)^2 \right] \;.
\end{equation*}
It can be easily verified that \eqref{S_h_def} can be written as
\begin{equation*}
\begin{split}
 S(\mathbf{h}_k) = \prod_{j = 1}^k \int_{- \infty}^{+ \infty}\left( \frac{f(Z_j | \rho_j + h_j)}{f(Z_j | \rho_j)} \right)^2 f(Z_j | \rho_j) d Z_j \;.
 \end{split}
\end{equation*}
Moreover, it can be noted that
\begin{equation*}
\begin{split}
\left( \frac{f(Z_j | \rho_j + h_j)}{f(Z_j | \rho_j)} \right)^2 =& \exp \left[ -2 \left( \phi_{\rho} (\rho_j + h_j) - \phi_{\rho}(\rho_j) \right) \right] \times \\
& \exp \left( 2 h_j \eta_{\rho}(Z_j) \right)
\end{split}
\end{equation*}
and therefore
\begin{equation*}
\begin{split}
\int_{- \infty}^{+ \infty}\left( \frac{f(Z_j | \rho_j + h_j)}{f(Z_j | \rho_j)} \right)^2 f(Z_j | \rho_j) d Z_j &= M^{-2}_{\eta_{\rho}} (h_j) \times \\
 \mathbb{E} & \left[ \exp \left( 2 h_j \eta_{\rho}(Z_j) \right) \right] \;.
\end{split}
\end{equation*}
Then, since
\begin{equation}
\mathbb{E} \left[ \exp \left( 2 h_j \eta_{\rho}(Z_j) \right) \right] = \exp \left( \phi_{\rho}(\rho_j + 2h_j) - \phi_{\rho}(\rho_j) \right)
\label{exp_2_h_eta_general}
\end{equation}
it can be easily seen that
\begin{equation*}
\begin{split}
\int_{- \infty}^{+ \infty}\left( \frac{f(Z_j | \rho_j + h_j)}{f(Z_j | \rho_j)} \right)^2 f(Z_j | \rho_j) d Z_j =& M^{-2}_{\eta_{\rho}} (h_j) M_{\eta_{\rho}} (2h_j)  \;,
\end{split}
\end{equation*}
thus getting
\begin{equation*}
S(\mathbf{h}_k) = \prod_{j=1}^{k} M^{-2}_{\eta_{\rho}} (h_j) M_{\eta_{\rho}} (2h_j) \;.
\end{equation*}
\section{MSE Expressions for ML Estimators}
\label{app:MSE_bounds_classical}
\subsection{Gaussian Distribution}
If the \label{app:MSE_bounds_classical_gauss}likelihood for $\xi$ is Gaussian distributed \eqref{example:gauss:pdf}, the MLE is given by \eqref{example:gauss:xi:ML}. Since the variance of the readings $U_j$ is $\sigma_{\xi}^2$ and the MLE \eqref{example:gauss:xi:ML} is unbiased, it is straightforward to see that
\begin{equation*}
\textrm{MSE} \left( \hat{\xi}_{\textrm{ML}} \right) = \frac{\sigma_{\xi}^2}{N} \;,
\end{equation*}
and similarly for $\hat{\psi}_{\textrm{ML}}$. Given \eqref{theta:ML} and Proposition \ref{prop:1}, it can be concluded that
\begin{equation}
\textrm{MSE} \left( \hat{\theta}_{\textrm{ML}} \right) = \frac{\sigma_{\xi}^2 + \sigma_{\psi}^2}{4N} \;.
\label{MSE_MLE_gaussian}
\end{equation}

\subsection{Exponential Distribution}
If the likelihood for $\xi$ is exponential distributed \eqref{example:exp:pdf}, the MLE is given by \eqref{ml:E}. Through simple algebra it can be seen that $U_{(1)}$ is exponentially distributed with parameter $\lambda_{\xi}^{'} = \lambda_{\xi} N$, so that $\textrm{Var} \left( \hat{\xi}_{\textrm{ML}} \right) = \frac{1}{\lambda_{\xi}^2 N^2}$. It can be noticed that $ \hat{\xi}_{\textrm{ML}}$ is a biased estimator for $\xi$, with bias $b_{\xi, \textrm{ML}} \overset{\Delta}{=}\frac{1}{ \lambda_{\xi} N}$. Similarly, $\textrm{Var} \left( \hat{\psi}_{\textrm{ML}} \right) = \frac{1}{\lambda_{\psi}^2 N^2}$ and $b_{\psi, \textrm{ML}} \overset{\Delta}{=}\frac{1}{ \lambda_{\psi} N}$. Therefore, given \eqref{theta:ML} and Proposition \ref{prop:1}, it can be concluded that
\begin{equation}
\begin{split}
\textrm{MSE}  \left( \hat{\theta}_{\textrm{ML}} \right) =
%& \frac{1}{4} \left( \textrm{Var}  \left( \hat{\xi}_{\textrm{ML}} \right) + \textrm{Var}  \left( %\hat{\psi}_{\textrm{ML}} \right) \right) + \\
%&  \frac{1}{4} \left( b_{\xi, \textrm{ML}} - b_{\psi, \textrm{ML}} \right)^2 = \\
& \frac{0.25}{N^2} \left( \frac{1}{\lambda_{\xi}^2} + \frac{1}{\lambda_{\psi}^2} \right) + \frac{0.25}{N^2} \left( \frac{1}{\lambda_{\xi}} - \frac{1}{\lambda_{\psi}} \right)^2 \;.
\end{split}
\label{MSE_MLE_exp_distr}
\end{equation}

\end{document}